\documentclass[12pt,tightenlines,eqsecnum,floats,showpacs,nofootinbib,amsmath,amssymb,aps,prd,superscriptaddress]{revtex4-1}
\usepackage{amsmath,amssymb}
\usepackage{amsfonts,amssymb,mathrsfs}
\usepackage[bookmarks=false,pdfstartview=FitH]{hyperref} % make sure it is the last of the packages

\def\be{\begin{equation}}
\def\ee{\end{equation}}
\def\nn{\nonumber}
\def\f{\frac}
\def\tf{\tfrac}
\def\sgn{{\rm sgn}}
\def\pl{{\rm Pl}}
\def\lp{\ell_\pl}
\def\mC{\mathcal{C}}
\def\mH{\mathcal{H}}

\def\mM{\mathcal{M}}

\def\mCHhom{\mC_H^{\rm hom}}
\def\mCHint{\mC_H^{\rm int}}

\def\b{\bar}
\def\d{\dot}
\def\h{\hat}

\def\v{\vec}
\def\wh{\widehat}

\def\dd{{\rm d}}
\def\na{\nabla}
\def\del{\partial}
\def\de{\delta}
\def\ga{\gamma}
\def\la{\lambda}

\def\vp{\varphi}
\def\oe{\mathring{e}}

\def\oq{\mathring{q}}
\def\bra{\langle}
\def\ket{\rangle}

\begin{document}

\title{Lattice Loop Quantum Cosmology: Scalar Perturbations}

\author{Edward Wilson-Ewing} \email{wilson-ewing@cpt.univ-mrs.fr}
\affiliation{Centre de Physique Th\'eorique, Aix-Marseille Univ, CNRS UMR 7332,
Univ Sud Toulon Var, 13288 Marseille Cedex 9, France}

\begin{abstract}

We study the scalar modes of linear perturbations in loop quantum cosmology.
This is done on a lattice where each cell is taken to be homogeneous
and isotropic and can be quantized via standard homogeneous loop
quantum cosmology techniques.  The appropriate interactions between
nearby cells are included in the Hamiltonian in order to obtain the correct
physics.  It is shown that the quantum theory is anomaly-free: the scalar
and diffeomorphism constraint operators weakly commute with the
Hamiltonian.  Finally, the effective theory encoding the leading order
quantum gravity corrections is derived and is shown to give the same
holonomy-corrected effective equations that have been obtained in
previous studies.

\end{abstract}

% Quantum Cosmology, Loop Quantum Gravity
% 98.80.Cq Field Theory Models of the Early Universe
% 04.60.Ds Canonical Quantization

\maketitle

\section{Introduction}
\label{s.intro}

Loop quantum cosmology (LQC) is a nonperturbative theory of quantum cosmology
motivated by loop quantum gravity.  Homogeneous and isotropic space-times
have been studied in considerable detail in LQC now and there is a growing
interest in the dynamics of linear perturbations as some effects specific to
LQC could potentially be detected in the cosmic microwave background or
in primordial gravitational waves.  This possibility is especially important
as cosmological observations are likely to provide the best testing ground
for theories of quantum gravity.

In order to study perturbations around a homogeneous background, it is of course
necessary to understand the background first.  This is under control as the LQC of
homogeneous and isotropic Friedmann-Lema\^itre-Robertson-Walker (FLRW) space-times
with a massless scalar field have been examined in the flat 
\cite{Ashtekar:2006wn, Ashtekar:2007em, MartinBenito:2009aj, Kaminski:2010yz},
closed \cite{Szulc:2006ep, Ashtekar:2006es}
and open cases \cite{Vandersloot:2006ws, Szulc:2007uk}, with a nonzero cosmological
constant \cite{Bentivegna:2008bg, Pawlowski:2011zf} and also with a Maxwell field
rather than the massless scalar field \cite{Pawlowski:2012}, in each case using the
so-called ``improved dynamics'' quantization.  Numerical and analytical
studies have shown that the
trajectory of sharply peaked states closely follows the classical trajectory predicted
by general relativity so long as the curvature is far from the Planck scale.  However,
deviations from general relativity appear in a contracting universe as the curvature
nears the Planck scale, at which point gravity becomes repulsive and causes a bounce.
In this way, LQC provides a quantum bridge between a classical contracting universe
at early times and a classical expanding universe at late times.

In addition, the dynamics of sharply peaked states are extremely well approximated
by what are called the effective dynamics.
The effective equations provide a quantum-corrected Hamiltonian constraint that can
be treated classically and the equations of motion include the modifications due to
quantum gravity effects.  For example, for the flat FLRW space-time the Friedmann
equation is modified to
\be H^2 = \f{8 \pi G}{3} \rho \left( 1 - \f{\rho}{\rho_c} \right), \ee
where the critical energy density is $\rho_c \cong 0.41 \rho_{\rm Pl}$
\cite{Taveras:2008ke}.  The effective equations have been studied for all of the
isotropic space-times considered so far in LQC, and in each case they have been
shown to provide an excellent approximation to the full quantum dynamics of sharply
peaked states, even in the presence of spatial curvature or with a nonzero
cosmological constant in which case the model is not exactly soluble.

Going beyond homogeneity and isotropy, the anisotropic Bianchi models
\cite{Ashtekar:2009vc, Ashtekar:2009um, WilsonEwing:2010rh} and the inhomogeneous
Gowdy model \cite{Garay:2010sk, MartinBenito:2010bh, MartinBenito:2010up} have
also been studied and, although they have not been studied in as much detail as
the FLRW models, it has been shown that the big-bang and big-crunch singularities
are resolved in these models as well.  The loop quantizations of the FLRW,
Bianchi and Gowdy space-times have recently been reviewed in some detail
in Refs.\ \cite{Ashtekar:2011ni, Banerjee:2011qu, Bojowald:2008zzb}.

As it is relatively simple to study quantum gravity effects in the effective
theory, it was in this setting that linear perturbations were first examined in
LQC.  The two main types of quantum geometry effects that appear in LQC (and loop
quantum gravity), inverse triad and holonomy effects, have been studied separately
in the effective treatments of linear perturbations.  Inverse volume effects
have been studied for tensor \cite{Bojowald:2007cd, Grain:2009eg}, vector
\cite{Bojowald:2007hv} and scalar perturbations \cite{Bojowald:2008gz,
Bojowald:2008jv}; and holonomy corrections have been determined for tensor
\cite{Bojowald:2007cd, Grain:2009eg}, vector \cite{Mielczarek:2011ph} and scalar
\cite{WilsonEwing:2011es, Cailleteau:2011kr, Cailleteau:2011mi} perturbations
as well.

These two different types of corrections are important in different settings: holonomy
corrections are relevant when the space-time curvature nears the Planck scale, while
inverse triad corrections become important when a physical length scale becomes
comparable to the Planck length $\lp$.  Some recent results suggest that the relevant
physical length scale is the wavelength of the inhomogeneous modes being studied,
in which case inverse triad effects vary from mode to mode, depending on the
wavelength of the mode \cite{WilsonEwing:2011es}.  (For a different point of view,
see \cite{Bojowald:2011iq} where it is instead suggested that it is the distance
between neighbouring nodes in the fundamental spin-network that is the relevant length
scale.)

Although work studying linear perturbations around FLRW space-times in the effective
theory has been very useful and has led to several interesting phenomenological
investigations (see e.g.\ \cite{Bojowald:2011iq, Copeland:2008kz, Grain:2009kw,
Grain:2010yv}), it is important to verify that there does indeed exist a quantum
theory whose dynamics, for sharply-peaked states, is well approximated by these
effective dynamics.  After all, the quantum theory should be the fundamental theory,
while the effective dynamics are merely a (very) useful approximation.  In this paper,
we will address this issue by constructing a quantum theory whose effective dynamics
are precisely those found in previous studies of holonomy-corrected equations for
scalar perturbations \cite{WilsonEwing:2011es, Cailleteau:2011kr, Cailleteau:2011mi}.
The quantum theory will also allow for investigations into the dynamics of states
that are not sharply peaked, something that is impossible in the effective theory.

Three methods have been proposed so far in order to model small perturbations
around an FLRW background.  The first is to use a lattice where each cell is
approximated by an FLRW space-time.  As the parameters of the FLRW model vary
from one cell to another, there are inhomogeneities at scales larger than the
volume of each cell.  If the parameters only vary slightly from one cell to the
next, then the perturbations are small and therefore this setup describes small
perturbations on an FLRW background.  This is similar to the separate universes
approach \cite{Wands:2000dp, Rigopoulos:2003ak} and was first suggested in the
context of LQC in \cite{Bojowald:2006qu} where the kinematics were studied, but
neither the Hamiltonian nor the scalar or diffeomorphism constraints were defined
in this work.  The second method is to treat the small perturbations as a test
quantum field on the LQC FLRW background, much as was done for a scalar field in
\cite{Ashtekar:2009mb}; a brief presentation of this programme can be found
in \cite{Agullo:2012ax}.  Finally, the third method, similar to the second,
is to perform a hybrid quantization where the perturbations are quantized
\`a la Fock on an LQC background.  This technique was pioneered for the study
of the the inhomogeneous Gowdy models \cite{Garay:2010sk, MartinBenito:2010bh,
MartinBenito:2010up} and has recently been used to model perturbations in
inflation \cite{FernandezMendez}.

In this paper, we will follow the first method.  We will build a cubical
lattice where each cell is taken to be homogeneous and isotropic, and there
will be some appropriate interactions between neighbouring cells in order to
obtain the correct physics.  The parameters of the homogeneous space-times
may only vary slightly from cell to cell, in order to ensure that one can
meaningfully speak of an average background, and small perturbations around
it.  This will require some restrictions on the initial conditions in order
to be sure that we are in the regime of small perturbations.  The kinematics
will be quite similar those defined in \cite{Bojowald:2006qu}, although
they will be presented in a slightly different way due to some recent
progress in homogeneous LQC \cite{Ashtekar:2006wn, Ashtekar:2007em}.  In
this paper, we will only consider scalar perturbations as this will
significantly simplify the model.  In Sec.\ \ref{s.dis}, we will briefly
comment upon a possible generalization which would allow vector and tensor
modes to be treated in a similar lattice LQC treatment.

The main limitation of the lattice LQC model that we shall introduce in this
paper is that it cannot be used to study perturbations with a wavelength $\la$
shorter than $\lp$.  The reason for this is the following: in order to have
a lattice that describes perturbations up to a given wavelength, an appropriate
number of cells is necessary.  However, for the homogeneous LQC model in each cell
to be trusted, the physical volume of each cell should remain much greater than
the Planck volume.  This means that it is only possible to describe perturbations
whose wavelengths remain much greater than $\lp$ at all times and therefore lattice
LQC cannot be used to describe cosmological models such as inflation where the
wavelength of the inhomogeneous modes seen today becomes smaller than $\lp$ at
early times.  Nonetheless, lattice LQC can be used in many other cosmological
scenarios, including the ekpyrotic and matter bounce models, where the wavelengths
of the relevant modes remain larger than the Planck length at all times.  Note
that an important consequence of this is that inverse triad corrections shall
always be negligible in lattice LQC as they only become important when
$\la / \lp \sim 1$ \cite{WilsonEwing:2011es}.

The outline of the paper is as follows.  In Sec.\ \ref{s.cl}, the classical
perturbation theory for scalar modes (in the appropriate variables for LQC)
is recalled from \cite{WilsonEwing:2011es}, and we present a discretization
of the theory.  In Sec.\ \ref{s.qm}, the restrictions necessary for wave
functions to represent small perturbation states are described, and then it
is shown that for these states the scalar, diffeomorphism and master constraint
operators weakly commute with the Hamiltonian.  The effective theory is
obtained in Sec.\ \ref{s.eff} and we conclude with a discussion in
Sec.\ \ref{s.dis}.

In this paper, we set $c=1$ but leave $G$ and $\hbar$ explicit so that it will
be easy to see whether contributions come from gravitational effects, quantum
effects or a combination of the two.  The Planck length is defined as
$\lp = \sqrt{G\hbar}$.  Finally, as the perturbations are linear, all terms
that are second order (or higher) in the perturbations can be dropped and
therefore all equations are understood to hold up to first order in perturbations
(with the sole exception of the Hamiltonian where second order terms are relevant
as they generate the dynamics of the linear perturbations).

\section{Cosmological Perturbations on a Lattice}
\label{s.cl}

In order to study small fluctuations around the flat FLRW cosmological background
(with topology $\mathbb{T}^3$), one allows small departures from homogeneity.  A
particularly nice coordinate choice for scalar perturbations is the longitudinal
gauge in which case the metric is given by (assuming zero anisotropic stress)
\be \dd s^2 = -(1 + 2 \psi) \dd t^2 + a^2 (1 - 2 \psi) \dd \v x^2, \ee
where $a(t)$ is the scale factor and depends only on time while $\psi(\v x, t)$
encodes the fluctuations away from the mean scale factor and varies both with
time and position.  We have chosen the line element so that the volume of the
3-torus, with respect to the background metric given by
$\dd \mathring{s}^2 = \dd \v x^2$, is 1.

The longitudinal gauge is a very useful choice as it can be viewed as the standard
FLRW metric, though with a scale factor that varies slightly with position.  This
property leads to drastic simplifications on the lattice where each cell is
approximated as being homogeneous.  In the longitudinal gauge, the homogeneous
metric in each cell is precisely the FLRW metric and therefore, the loop quantization
of each cell will be relatively straightforward as the LQC of flat FLRW space-times
is well known \cite{Ashtekar:2006wn, Ashtekar:2007em}.  It is this major
simplification that motivates the choice of the longitudinal gauge here.  In
other gauges for scalar perturbations (and for vector and tensor modes no matter
the gauge), the metric is necessarily nondiagonal and then the well understood
results of isotropic LQC cannot be applied in a lattice LQC setting.

In the first part of this section we will briefly present a canonical treatment of
scalar cosmological perturbations in the longitudinal gauge, closely following the
results of \cite{WilsonEwing:2011es}.  As we are only considering linear perturbations,
the scalar, diffeomorphism and Gauss constraints can be truncated to first order.
(In fact, the Gauss constraint vanishes to first order due to the choice of the
longitudinal gauge and it can therefore be ignored.)  However, in the Hamiltonian,
terms that are second order in the perturbations generate the dynamics of first order
perturbations and therefore second order terms are relevant for the Hamiltonian.
Note that the Hamiltonian vanishes to first order, but not to second order. See
\cite{Mukhanov:1990me, Langlois:1994ec} for further discussions regarding these
issues. The main ingredients introduced here will therefore be the Hamiltonian, to
second order in perturbation theory, and the scalar and diffeomorphism constraints,
to first order in perturbation theory.  This will be done using variables that will
be convenient later for the nonperturbative loop quantization of the FLRW space-times
in each cell of the lattice. This review will be brief, for further details concerning
the canonical treatment of cosmological perturbations see \cite{WilsonEwing:2011es,
Mukhanov:1990me, Langlois:1994ec}.

In the second part of this section, we will discretize the Hamiltonian as well as
the scalar and diffeomorphism constraints on a cubical lattice.  We will see that
the Hamiltonian naturally breaks into a homogeneous term and an interaction term;
as we shall see, this split will allow us to perform a loop quantization on a
lattice in Sec.\ \ref{s.qm}.

\subsection{Perturbations in Tetrad Variables}
\label{ss.cont}

In order to study the perturbations in the loop gravity framework, it is
necessary to use the Ashtekar connection and densitized triads as our basic variables.
Since the metric is diagonal, we can parametrize the densitized triads $E^a_i = \sqrt{|q|}
e^a_i$ by
\be E^a_i = p \sqrt{\oq} \oe^a_i, \qquad {\rm where} \qquad p = a^2 (1 - 2 \psi), \ee
where $p$ is a function of position and time.  We will drop the arguments for the remainder
of Sec.\ \ref{ss.cont} in order to simplify the notation, except in the definition of the
Poisson brackets where they are essential.  We assume that $\int_\mM \psi = 0$ in order
to ensure that $a$ is the true background.  Therefore, $\int_\mM p = a^2$.

The Ashtekar connection is given by
$A_a^i = \Gamma_a^i + \ga K_a^i$, where $\ga$ is the Barbero-Immirzi parameter, $\Gamma_a^i$
is the spin-connection and $K_a^i = K_{ab} e^{bi}$ is related to the extrinsic curvature.
A very useful propery of the Ashtekar connection in this context is that its diagonal
terms solely come from $K_a^i$, while its off-diagonal terms solely come from $\Gamma_a^i$.
As the densitized triads are diagonal, only the diagonal part of the
Ashtekar connection will appear in the induced symplectic structure and therefore it is
convenient to parametrize the Ashtekar connection by
\begin{align} A_x &= c \, \tau_1 - (\del_z \psi) \, \tau_2 + (\del_y \psi) \, \tau_3, \nn \\
A_y &= (\del_z \psi) \, \tau_1 + c \, \tau_2 - (\del_x \psi) \, \tau_3, \\
A_z &= - (\del_y \psi) \, \tau_1 + (\del_x \psi) \, \tau_2 + c \, \tau_3, \nn \end{align}
where $A_a = A_a^i \tau_i$ and the $\tau_i$ are a basis of the Lie algebra of SU(2) such
that $\tau_i \tau_j = \tf{1}{2} \epsilon_{ij}{}^k \tau_k - \tf{1}{4} \de_{ij} \mathbb{I}$
and we have solved for the spin-connection in terms of the perturbation $\psi$.
Following this definition, we find that the induced symplectic structure on our phase
space gives the following nonzero Poisson bracket:
\be \{ c(\v x), p(\v y) \} = \f{8 \pi G \ga}{3} \de^3(\v x - \v y). \ee
We will consider the case where the matter field is given by a massless scalar field in
which case the fundamental Poisson bracket for the matter sector is given by
\be \label{poiss1} \{ \vp(\v x), \pi_\vp(\v y) \} = \de^3(\v x - \v y). \ee

In order to obtain the Hamiltonian constraint, one must first derive the scalar,
diffeomorphism and Gauss constraints.  For linear perturbations around the flat FLRW
space-time in the longitudinal gauge, we find that the Gauss constraint is automatically
satisfied, while the scalar constraint to second order in perturbation theory is
\be \mH = \sqrt{\oq} \Bigg[\f{-\sqrt{|p|}}{8 \pi G} \bigg( \f{3 c^2}{\ga^2}
+ 2 \Delta \psi - \left( \v\na \psi \right)^2 \bigg) + \f{\pi_\vp^2}{2 p^{3/2} \oq}
+ \f{\sqrt{p}}{2} \left( \v\na \vp \right)^2 \Bigg]
\approx 0, \ee
where $\v\na$ is the gradient and $\Delta$ denotes the Laplacian.  The diffeomorphism
constraint to first order in perturbation theory%
\footnote{It is only necessary to calculate the diffeomorphism constraint to first order
as the shift vector $N^a$ is necessarily small and as the diffeomorphism constraint
appears in the Hamiltonian as $N^a \mH_a$, terms that are first order in the diffeomorphism
constraint will provide second order contributions to the Hamiltonian.}
% end footnote
is given by
\be \mH_a = \f{\sqrt{\oq} p}{4 \pi G \ga} \bigg[ \del_a c + c \, \del_a \psi \bigg]
+ \pi_\vp \del_a \vp \approx 0. \ee
Finally, since $N = 1 + \psi$ and $N^a = 0$ in the longitudinal
gauge, the Hamiltonian constraint is given by
\be \mC_H = \int_\mM \bigg( N \mH + N^a \mH_a + \Lambda^i \mathcal{G}_i \bigg) =
\int_\mM \bigg( 1 + \psi \bigg) \mH.
\ee
See \cite{WilsonEwing:2011es} for a derivation of the results presented above.	

We will now introduce a new pair of variables which are convenient for the loop quantization,
\be \nu = \f{\sgn(p) |p|^{3/2}}{2 \pi \ga \lp^2} = \f{a^3(1 - 3\psi)}{2 \pi \ga \lp^2}, \ee
and
\be b = \f{c}{\sqrt{|p|}} = \f{c}{(2 \pi \ga \lp^2)^{1/3} |\nu|^{1/3}}, \ee
which are canonically conjugate,
\be \label{poiss2} \{ b(\v x), \nu(\v y) \} = \f{2}{\hbar} \, \de^3(\v x - \v y). \ee

In these variables, the Hamiltonian constraint is given by
\be \mC_H = \mCHhom + \mCHint, \ee
where $\mCHhom$ is the ultralocal (or ``homogeneous'') portion of the scalar constraint which
does not include any derivatives and is given by

\be \mCHhom = \int_\mM \sqrt{\oq} \, (1 + \psi) \left[ -\f{3 \hbar}{4 \ga} |\nu| b^2
+ \f{1}{4 \pi \ga \lp^2 |\nu|} \f{\pi_\vp^2}{\oq} \right], \ee
and the interaction terms are
\begin{align} \mCHint &=  \int_\mM \sqrt{\oq} \, \f{(2 \pi \ga \lp^2)^{1/3} |\b\nu|^{1/3}}{8 \pi G}
\left[ -2 \Delta \psi + \left(\v\nabla \psi \right)^2
+ 4 \pi G \left( \v\nabla \vp \right)^2 \right] \nn \\
&= \int_\mM \sqrt{\oq} \f{(2 \pi \ga \lp^2)^{1/3} |\nu|^{1/3}}{8 \pi G}
\left[ \left(\v\nabla \psi \right)^2 + 4 \pi G \left( \v\nabla \vp \right)^2 \right], \end{align}
where on the first line we introduced
\be \label{cl-bnu} \b\nu = \int_\mM \sqrt{\oq} \, \nu, \ee
and used the relation $(1 + \psi) \nu^{1/3} = (1 + \psi) \b\nu^{1/3} (1 - \psi) = \b\nu^{1/3}$ which
holds to first order in perturbation theory.  Also, the first term on the first line vanishes as
$\int_\mM \nabla^2 \psi = 0$ (note that $\b\nu$ can be pulled out of the integral).  In the Hamiltonian
constraint, terms that are first and second order in the perturbations are relevant for the dynamics,
while higher order terms can be ignored.

Note that the homogeneous term in the scalar constraint here has the exact form of the scalar constraint
of the flat homogeneous and isotropic FLRW space-time obtained in \cite{Ashtekar:2007em} (with the
difference that now $b$ and $\nu$ are position-dependent.)  We will take advantage of this in the
quantization procedure in the following section.

This Hamiltonian, expressed in terms of the basic variables $b, \nu, \vp, \pi_\vp$, provides the dynamics.
It follows that all of the occurences of $\psi$ in $\mC_H$ must be replaced by
\be \label{cl-psi} \psi = \f{\b\nu - \nu}{3 \b\nu}. \ee
Note that this means that the nonlocal quantity $\b\nu$ will appear in the Hamiltonian constraint and
therefore, in principle, the equations of motion could be nonlocal.  However, one can check that all
of the nonlocal terms vanish and that the resulting equations of motion are local, as expected.

The equations of motion are obtained from the Poisson bracket via
\be \d{\mathcal{O}}(\v x) = \{ \mathcal{O}(\v x), \mC_H \}, \ee
and the equations of motion for the basic variables, truncated to first order in perturbation theory
and only considering positive values of $\nu$ for simplicity, are given by
\begin{align} \label{cl-eom1} \d\vp &= \f{1 + \psi}{2 \pi \ga \lp^2 \nu} \f{\pi_\vp}{\sqrt{\oq}}, \\
\label{cl-eom2} \d\pi_\vp &= \left( 2 \pi \ga \lp^2 \nu \right)^{1/3} \Delta \vp, \\
\label{cl-eom3} \d\nu &= \f{3(1 + \psi)}{\ga} \nu b, \\
\label{cl-eom4} \d b &= -\f{3(1 + \psi)}{2\ga} b^2 - \f{1 + \psi}{2 \pi \ga \hbar \lp^2 \nu^2}
\f{\pi_\vp^2}{\oq}. \end{align}
In order to obtain the last equation of motion, the relations $\mH = 0$ and $\int_\mM \nabla^2 \psi = 0$
are needed.

Finally, the initial data is constrained by $\mH$ and $\mH_a$, truncated to first order in the perturbations:
\be \mH = -\f{3 \hbar}{4 \ga} |\nu| b^2 + \f{\pi_\vp^2}{4 \pi \ga \lp^2 |\nu|}
- \f{(2 \pi \ga \lp^2 |\b\nu|)^{1/3}}{4 \pi G} \Delta \psi \approx 0, \ee
\be \mH_a = \f{\hbar \nu}{2} \del_a b + \pi_\vp \del_a \vp \approx 0. \ee
A nice feature of these variables is the particularly simple form of the diffeomorphism constraint.
One can check that the scalar and diffeomorphism constraints weakly commute with the Hamiltonian,
showing that the dynamics preserves the contraint surface.

This concludes the presentation of the ingredients necessary for a canonical treatment of
cosmological perturbations in loop variables.

\subsection{Discretization}
\label{ss.latt}

To date, it is only known how to quantize homogeneous models in LQC (aside from hybrid
quantization schemes).  Therefore, it seems that the simplest way to incorporate
inhomogeneities in LQC is to introduce a lattice where each cell in the lattice is
taken to be homogeneous but the gravitational field and the scalar field are allowed
to vary from cell to cell, thus allowing inhomogeneous degrees of freedom at scales
larger than the size of the cells.  By following this procedure, we can use the
techniques developed for homogeneous models for LQC in order to quantize each cell
and thus obtain a quantum theory which allows large-scale inhomogeneities.

For simplicity, we decompose the space into a cubical lattice with $Z^3$ cells and we
label each cell by a vector $\v z = (z_1, z_2, z_3)$ where the $z_a$ are integers
running from 1 to $Z$.  We also introduce the unit cell displacement vectors $\h z_a$
where, e.g., $\h z_1 = (1, 0, 0)$.  Note that the distance between the centres of two
neighbouring cells with respect to the fiducial metric is $1/Z$ and therefore the
norm of $\h z_a$ with respect to the fiducial metric is $1/Z$.

The Hamiltonian formalism introduced in the previous section can be defined on the
lattice in a relatively straightforward procedure.  First, the induced symplectic
structure has changed to
\be \label{poisson1} \{ b(\v z_1), \nu(\v z_2) \} = \f{2 Z^3}{\hbar} \de_{\v z_1, \v z_2}, \ee
\be \label{poisson2} \{ \vp(\v z_1), \pi_\vp (\v z_2) \} = Z^3 \de_{\v z_1, \v z_2}, \ee
where $\de_{\v z_1, \v z_2}$ is the three-dimensional Kronecker delta.  It would be possible
to rescale $\nu \to Z^3 \nu$ and $\pi_\vp \to Z^3 \pi_\vp$ in order to remove the presence
of the $Z^3$ in the Poisson brackets, but we will not do this in order to explicitly
show the presence of the lattice.

Now, in order to implement the scalar and diffeomorphism constraints on the lattice, the
derivatives must be approximated by differences.  There is a choice to be made concerning
the exact form of the differences used and we will choose
\be \label{def-disc-deriv} \del_a f(\v x)\Big|_{\v z}
\cong \f{Z}{3} \Big( f(\v z + 2 \h z_a) - f(\v z - \h z_a) \Big)
\equiv (\nabla_\dd)_a f(\v z), \ee
\begin{align} \left( \v \nabla f(\v x) \right)^2\Big|_{\v z} &
\cong \f{Z^2}{9} \sum_{a=1}^3 \Big( f(\v z + 2 \h z_a) - f(\v z - \h z_a) \Big)
\Big( f(\v z + \h z_a) - f(\v z - 2 \h z_a) \Big) \nn \\ &
\equiv \left( \v \nabla f(\v z) \right)_\dd^2, \label{def-disc-diff}
\end{align}
\begin{align} \Delta f(\v x)\Big|_{\v z}
\cong \f{Z^2}{18} \sum_{a=1}^3 \Big( & f(\v z + 4 \h z_a) + f(\v z + 2 \h z_a)
- 2 f(\v z + \h z_a) \nn \\ & \label{def-disc-lapl}
- 2 f(\v z - \h z_a) + f(\v z - 2 \h z_a) + f(\v z - 4 \h z_a) \Big)
\equiv \Delta_\dd f(\v z). \end{align}
This choice is a good one because the discretization of $(\v \nabla f)^2$ is such that
there are no $f(\v z)^2$ terms which can cause anomalies in the quantum theory (instead
there are terms like $f(\v z + \h z_a)f(\v z - \h z_a)$ which do not cause anomalies).
Note that simpler discretizations are possible classically, but become problematic
in the quantum theory.  The discretization $\Delta_\dd f(\v z)$ has been chosen as it
is the one which is compatible with $(\v \nabla f)_\dd^2$,
\be \f{\de}{\de f(\v z)} \sum_{\v y} \left( \v \nabla f(\v y) \right)_\dd^2
= - 2 \Delta_\dd f(\v z), \ee
and similarly $(\nabla_\dd)_a f$ is compatible with $\Delta_\dd f$ as
\begin{align} \Delta_\dd f(\v z) &= \f{Z}{6} \sum_{a=1}^3 \Big(
(\nabla_\dd)_a f(\v z + 2 \h z_a) + (\nabla_\dd)_a f(\v z)
- (\nabla_\dd)_a f(\v z - \h z_a) - (\nabla_\dd)_a f(\v z - 3 \h z_a) \Big) \nn \\
&= \f{1}{2} \sum_{a=1}^3 \Big( (\nabla_\dd)_a [(\nabla_\dd)_a f](\v z)
+ (\nabla_\dd)_a [(\nabla_\dd)_a f](\v z - 2 \h z_a) \Big). \end{align}
Note also that $(\v \nabla f)_\dd^2$ can be written in terms of $(\nabla_\dd)_a f$.

As the discrete difference operators approximate derivatives, the product and polynomial
rules should hold for the difference operators,
\be \label{l-prod} (\nabla_\dd)_a [fg] \cong f (\nabla_\dd)_a [g] + g (\nabla_\dd)_a [f], \ee
\be \label{l-poly} (\nabla_\dd)_a [f^n] \cong n f^{n-1} (\nabla_\dd)_a [f]. \ee
Of course, these relations are not exact as we are working on a lattice, but they are
valid within the approximation scheme we are using; they can be derived if the
discrete derivative \eqref{def-disc-deriv} is assumed to be small, as it should be
in the case of linear perturbations on a homogeneous background.  This derivation is
done explicitly in the quantum setting in Sec.\ \ref{ss.states}.

Finally, the discrete versions of the relations \eqref{cl-bnu} and \eqref{cl-psi},
\be \b\nu = \f{1}{Z^3} \sum_{\v z} \nu(\v z) \qquad {\rm and} \qquad
\psi(\v z) = \f{1}{3} - \f{\nu(\v z)}{3 \b\nu}, \ee
will be needed, as well as the condition that $\sum_{\v z} \psi(\v z) = 0$ which
ensures that $\b\nu$ captures the homogeneous mode of $\nu$.

It is now possible to write the Hamiltonian constraint for the lattice.  The homogeneous
part of the Hamiltonian becomes
\be \label{d-chom}
\mCHhom = \f{1}{Z^3} \sum_{\v z} \left(\f{4}{3} - \f{\nu(\v z)}{3 \b \nu} \right)
\left[ - \f{3 \hbar}{4 \ga} |\nu(\v z)| b(\v z)^2
+ \f{\pi_\vp(\v z)^2}{4 \pi \ga \lp^2 |\nu(\v z)|} \right],
\ee
and the interaction terms are
\be \label{d-cint}
\mCHint = \f{1}{Z^3} \sum_{\v z} \f{\left(2 \pi \ga \lp^2 |\nu(\v z)| \right)^{1/3}}
{8 \pi G} \bigg[ \f{1}{9 \nu(\v z)^2} \Big( \v \nabla \nu(\v z) \Big)_\dd^2
+ 4 \pi G \Big( \v \nabla \vp(\v z) \Big)_\dd^2 \bigg]. \ee
Similarly, the scalar constraint on the lattice is given by
\be \label{d-scal}
\mH(\v z) = -\f{3 \hbar}{4 \ga} |\nu(\v z)| b(\v z)^2
+ \f{\pi_\vp(\v z)^2}{4 \pi \ga \lp^2 |\nu(\v z)|}
+ \f{\left(2 \pi \ga \lp^2 \right)^{1/3}}{12 \pi G |\nu(\v z)|^{2/3}}
\Delta_\dd |\nu(\v z)|, \ee
and the diffeomorphism constraint is
\be \label{d-diff}
\mH_a(\v z) = \f{\hbar \nu(\v z)}{2} (\nabla_\dd)_a b(\v z)
+ \pi_\vp(\v z) (\nabla_\dd)_a \vp(\v z). \ee

The equations of motion can be obtained from $\mC_H = \mCHhom + \mCHint$ and it is easy to check
that they are the discretized equivalent of Eqs.\ \eqref{cl-eom1}---\eqref{cl-eom4}.

It is also possible to determine how the scalar and diffeomorphism constraints evolve with time on
the lattice:
\begin{align} \d \mH(\v z) &= \f{1}{(2 \pi \ga \lp^2 \nu(\v z))^{2/3}}
\left[ \f{\hbar}{2} \nu(\v z) \Delta_\dd b(\v z)
+ \pi_\vp (\v z) \Delta_\dd \vp(\v z) \right] \nn \\ &
= \f{1}{2 (2 \pi \ga \lp^2 \nu(\v z))^{2/3}} \sum_{a=1}^3
\Big[ (\nabla_\dd)_a \mH_a(\v z)
+ (\nabla_\dd)_a \mH_a(\v z - 2 \h z_a) \Big] \approx 0, \end{align}
\be \d \mH_a(\v z) = -\f{1}{3 \nu(\v z)} \mH(\v z) (\nabla_\dd)_a \nu(\v z)
\approx 0, \ee
where we have used the relation $(\nabla_\dd)_a \psi = -[(\nabla_\dd)_a \nu]/3\nu$
(note that analogous relations exist for each of the discrete derivative operators),
and the product and polynomial rules \eqref{l-prod} and \eqref{l-poly}.

This shows that the scalar and diffeomorphism constraints are preserved by $\mC_H$,
up to some small errors introduced by terms that are second order or higher in the
perturbations and can therefore be ignored.

The most important result of this section is that the Hamiltonian can be seen as a
flat FLRW space-time in each cell (although the lapse has an unusual form), with
$b, \nu, \vp, \pi_\vp$ varying from cell to cell, plus interactions between cells.
This will allow us to quantize this model in the next section by performing
the standard LQC quantization for FLRW space-times in each cell, and then work with
some interaction terms in the Hamiltonian that are relatively easy to handle.

\section{Lattice Loop Quantum Cosmology}
\label{s.qm}

In this section, we will quantize the discretized model of linear perturbations on a flat
FLRW background which was presented in the previous section.  Since in the discrete
version of the theory the metric in each cell is FLRW, it will be possible to quantize
the discretized theory cell by cell using methods that were first developed for isotropic
spaces in LQC.

However, there will be one important difference: in standard LQC, the scalar constraint vanishes
exactly and the Dirac quantization procedure for constrained systems is implemented exactly.
This is not possible in lattice LQC as even in the classical theory the scalar and diffeomorphism
constraints only vanish to first order, as does the Hamiltonian.  Because of this, in the quantum
theory it is only possible to demand that the constraints \emph{approximately} annihilate physical
states.  To be precise, the amplitude $A^2$ of the linear perturbations of a wave function $\Psi$
is determined by demanding that the norm of any relevant linear perturbation operator acting on
$\Psi$ is (at most) of the order of $A^2 \ll 1$.  Then, for the constraint $\h\mC$ to hold to
first order of perturbation theory, the norm of $\h \mC \Psi$ must be of the order of $A^4$.

Also, in lattice LQC there is more than one constraint and therefore the constraint algebra is
not trivial: it is necessary to check that the scalar and diffeomorphism constraints commute
weakly with the Hamiltonian, to linear order (i.e., the norm of the terms that are not weakly
zero acting on the wave function must be of the order of $A^4$ or smaller).  Although there are
some nontrivial checks to be performed in order to ensure that the quantum dynamics preserve the
constraints, we do not have access to a full constraint algebra here.  This is because we are
working in the longitudinal gauge and the lapse and shift have been fixed in order to preserve
the gauge.  Therefore, as one cannot freely choose any lapse or shift, the full constraint
algebra is not available.  Nonetheless, the commutators between the scalar and diffeomorphism
constraints with the Hamiltonian are an important check and show that lattice LQC is anomaly
free.

In this section, we will begin by defining the kinematical Hilbert space for the lattice, first cell
by cell and then for the entire lattice.  We will next discuss the conditions states must satisfy
in order to represent small perturbations, before constructing the Hamiltonian constraint operator
which will provide the dynamics and define the physical Hilbert space.  Finally, we will show
that the scalar and diffeomorphism constraint operators weakly commute with the Hamiltonian.

\subsection{The Kinematical Hilbert Space}
\label{ss.kincell}

The kinematical Hilbert space of lattice LQC is the tensor product of the kinematical Hilbert
spaces for each cell in the lattice.  In turn, the kinematical Hilbert space of each cell is
the tensor product of the gravitational and matter kinematical Hilbert spaces.

We shall briefly recall the kinematical Hilbert space for the gravitational sector in homogeneous
and isotropic LQC for the sake of completeness.  The kinematical Hilbert space is usually
presented in the $p$ representation where the operators corresponding to fluxes of the densitized
triads and holonomies of the connection are defined.  However, in the loop quantization, it turns
out that the holonomies appearing in the Hamiltonian are those with a length of
$\b\mu = \la/\sqrt{|p|}$ which depends on the value of $p$.  (Here $\la$ is the square root of the
minimal nonzero eigenvalue of the area operator in loop quantum gravity and has dimensions of length.)
Because of this nontrivial dependence on $p$, it is useful to work in a different representation
where the fundamental operators correspond to $\nu$ and complex exponentials of $b$.  For more
details of this, see e.g.\ \cite{Ashtekar:2007em}.  It is the kinematical Hilbert space in this
representation that will be presented here.

In the $| \nu \ket$ representation, the gravitational kinematical Hilbert space (for one of
the cells in the lattice) is spanned by wave functions $\Psi(\nu)$ that are a countable linear
combination of the basis states,
\be | \Psi \ket = \sum_\nu \Psi(\nu) | \nu \ket, \quad {\rm where} \quad \sum_\nu |\Psi(\nu)|^2
< \infty. \ee
Note that this is a discrete sum, not an integral.  The inner product is given by
\be \label{innerp} \bra \nu' | \nu \ket = \delta_{\nu', \nu}, \ee
where $\delta_{\nu', \nu}$ is a Kronecker delta rather than a Dirac delta.

As the orientation of the triads (right- or left-handed) is encoded in the sign of $\nu$, the parity
operator $\h \Pi$ which flips the orientation of the physical triads acts as
\be \label{parity} \h \Pi \Psi(\nu) = \Psi(-\nu). \ee

Following loop quantum gravity, the operators which are well-defined in LQC are holonomies and areas.
Due to the homogeneous and isotropic nature of each cell, it suffices to consider holonomies which are
parallel to the sides of the cell, and the physical areas of the sides of the cell.  However, as
mentioned above, since the holonomies appearing in the LQC Hamiltonian are of the form of
$\exp[ \pm i (\la / \sqrt{|p|}) c]$ \cite{Ashtekar:2006wn, Ashtekar:2007em}, it is useful to change
variables to $b$ and $\nu$.  Then the first fundamental operator that is to be defined is $\h\nu$ which
acts by multiplication,
\be \h \nu | \nu \ket = \nu | \nu \ket, \ee
and is related to the volume (of a cell),
\be \label{vol-op} \wh{V} | \nu \ket = 2 \pi \ga \lp^2 \, \h{\nu} / Z^3 | \nu \ket. \ee
The other fundamental operator is the holonomy operator that acts by shifting $\nu$,
\be \wh{\exp (i \ell b) } | \nu \ket = | \nu - 2 Z^3 \ell \ket. \ee

The kinematical Hilbert space for the wave functions $\Psi(\vp)$ describing the matter sector
is the space of square-integrable functions $L^2(\mathbb{R}, \dd \vp)$.  The basic operators
are defined as usual: $\h \vp$ acts by multiplication while
$\h \pi_\vp = -i \hbar \dd / \dd \vp$ acts by differentiation.

The complete kinematical Hilbert space of one cell is given by the tensor product of
the gravitational and matter Hilbert spaces, while the kinematical Hilbert space of the entire
space is the tensor product of the Hilbert spaces of each cell.

As the physics is invariant under a change of the overall orientation of the triads, the complete
state should be invariant under the parity transformation
\be \label{tot-parity}
\wh{\Pi_{\rm tot}} \Psi(\nu_1, \nu_2, \ldots, \nu_{Z^3})
= \Psi(-\nu_1, -\nu_2, \ldots, -\nu_{Z^3}). \ee
Note however that a typical state will not be invariant under the action of a single $\h \Pi(\v z)$
as this only reverses the orientation in one cell.

Finally, it is useful to define the operator corresponding to the mean value of $\h \nu(z_i)$.
The operator $\h{\b \nu}$ is simply given by
\be \h{\b \nu} \, \Psi(\nu_1, \nu_2, \ldots, \nu_{Z^3}) = \f{1}{Z^3} \sum_{i=1}^{Z^3} \nu_i \,
\Psi(\nu_1, \nu_2, \ldots, \nu_{Z^3}). \ee
Similar ``barred'' operators can be defined for other operators by averaging over all of the cells
in the lattice, but the $\h{\b \nu}$ operator is the only ``barred'' operator that appears in the
Hamiltonian.

\subsection{Restrictions on the States}
\label{ss.states}

Since we are interested in studying small perturbations, it is necessary to restrict our attention
to states where the wave function in each cell only varies slightly from one cell to another.  We
will impose two conditions in order to ensure that the wave functions describe small perturbations
away from homogeneity.

The first condition is that the parity cannot vary from one cell to another.  Therefore, the wave
functions may only have support on the two configurations where the sign of the arguments of the wave
function are all equal, i.e., $\Psi(|\nu_1|, |\nu_2|, \ldots, |\nu_{Z^3}|)$ and $\Psi(-|\nu_1|, -|\nu_2|, \ldots,
-|\nu_{Z^3}|)$.  It follows that%
\footnote{This can only be done if the evolution generated by the Hamiltonian operator does not couple
the positive and negative arguments of the wave function.  As we shall see, the Hamiltonian satisfies
this property.}
% end footnote
%
\be \label{parity-cond} \Psi(\ldots, |\nu_i|, \ldots, -|\nu_j|, \ldots) = 0, \qquad \forall \: i, j. \ee
Since the wave functions only have support on the sections where the $\nu_i$ are either all positive or
all negative, and these two sections are related by the parity transformation
\be \Psi(-|\nu_1|, -|\nu_2|, \ldots, -|\nu_{Z^3}|) = \wh{\Pi_{\rm tot}}
\Psi(|\nu_1|, |\nu_2|, \ldots, |\nu_{Z^3}|), \ee
which must be conserved by the dynamics of the system, we can restrict our attention to the ``positive''
portion of the wave function where all of the arguments satisfy the condition $\nu_i \ge 0$.  This is
what we will do for the remainder of the paper.

The second condition is that the perturbations in both the gravitational and matter fields must be small
and therefore the difference between an operator acting on different cells should be small as this is,
by construction, a measure of the perturbations around the homogeneous background.  To be more precise,
we ask that the norm of a wave function acted on by some difference operator $\h A(\v z_1) - \h A(\v z_2)$
be much smaller than the same wave function acted on by the operator $\h A(\v z_3)$, for all
$z_1, z_2, z_3$.  We can write this condition on the wave function as
\be \label{small-diff} \left( \h A(\v z_1) - \h A(\v z_2) \right) \Psi
\ll \h A(\v z_3) \: \Psi, \ee
where `$\ll$' indicates that the norm of the resulting wave function on the left hand side is much
smaller than on right hand side.  Similarly, we also have
\be \label{second-order}
\left( \h A(\v z_1) - \h A(\v z_2) \right) \left( \h B(\v z_3) - \h B(\v z_4) \right) \Psi
\ll \left( \h A(\v z_5) - \h A(\v z_6) \right) \: \Psi. \ee
This last relation holds regardless of the ordering of the two difference operators on the left hand
side.  An important point is that these relations must hold whether the $\h A$ and $\h B$ operators
act on the gravitational or matter parts of the wave function.

As we are only interested in linear perturbations, we will drop all terms that are second order
in terms of difference operators, i.e.\ terms like the one on the left hand side of the Eq.\
\eqref{second-order} above, except in the Hamiltonian constraint operator where they generate
nontrivial dynamics for the linear perturbations.

This last condition implies several other useful (approximate) relations that hold so long as we
work with linear perturbations.  The first one is simply
\begin{align}
\h A(z_1) \left( \h B(z_3) - \h B(z_4) \right) \Psi
&= \left( \h A(z_1) - \h A(z_2) \right) \left( \h B(z_3) - \h B(z_4) \right) \Psi \nn \\ &
\qquad + \h A(z_2) \left( \h B(z_3) - \h B(z_4) \right) \Psi, \nn \\
\label{q-bar}
& \cong \h A(z_2) \left( \h B(z_3) - \h B(z_4) \right) \Psi,
\end{align}
where we see that no matter which cell the operator $\h A$ acts on, the result will be the same
up to the first order in perturbations.  An important point is that this relation holds
whether $\h A$ is on the right or the left side of the difference operator.  Note also that
the prefactor to Eq.\ \eqref{q-bar} could be chosen to be the barred ``average'' operator
$\h{\b A}$ instead of $\h A(z_2)$.

Another useful relation is the quantum equivalent to the product rule that we introduced for
the lattice in Eq.\ \eqref{l-prod}:
\begin{align} 
\left( (\h A \h B)(z_1) - (\h A \h B)(z_2) \right) \Psi
&= \left( \h A(z_1) - \h A(z_2) \right) \h B(z_2) \Psi
+ \h A(z_1) \left( \h B(z_1) - \h B(z_2) \right) \Psi \nn \\ &
\label{q-prod}
\cong \left( \h A(z_1) - \h A(z_2) \right) \h B(z_3) \Psi
+ \h A(z_4) \left( \h B(z_1) - \h B(z_2) \right) \Psi,
\end{align}
where on the second line we used the relation \eqref{q-bar} twice.  Note that this relation
follows from the condition \eqref{second-order}, it is not a further assumption.  This last
relation immediately implies a relation analogous to the polynomial derivative,
\be \label{q-poly} \left( \h A(z_2)^n - \h A(z_3)^n \right) \Psi
\cong n \h A(z_1)^{n-1} \left( \h A(z_2) - \h A(z_3) \right) \Psi, \ee
which holds for fractional and negative $n$ so long as $\h A$ is invertible.  The existence
of these relations for difference operators ---that are very similar to those for differential
operators--- should not be surprising as we are only interested in linear perturbations
and we are discarding all higher order terms.

The final relation we will present here is obtained by
\begin{align}
(\h A \h B)(z_1) \left( \h C(z_3) - \h C(z_4) \right) \Psi
& \cong \h A(z_1) \h B(z_2) \left( \h C(z_3) - \h C(z_4) \right) \Psi \nn \\
& = \h B(z_2) \h A(z_1) \left( \h C(z_3) - \h C(z_4) \right) \Psi \nn \\
& \cong (\h B \h A)(z_2) \left( \h C(z_3) - \h C(z_4) \right) \Psi \nn \\
\label{drop-comm}
& \cong (\h B \h A)(z_1) \left( \h C(z_3) - \h C(z_4) \right) \Psi,
\end{align}
where we have used the relation \eqref{q-bar} repeatedly and, on the second
line, freely commuted the $\h A$ and $\h B$ operators as they act on
different cells (this requires the assumption that both of the operators act
only on one cell).  What this calculation shows is that terms of the form
\be [ \h A, \h B] (z_1) \left( \h C(z_3) - \h C(z_4) \right) \Psi, \ee
even though they may not look as though they are second order in perturbations,
can in fact be neglected.  This relation is equivalent to
\be \label{small-comm} [ \h A, \h B] (z) \, \Psi \ll (\h A \h B)(z) \, \Psi, \ee
which, although it at first seems to be quite different from \eqref{second-order},
is in fact implied by it.

A consequence of the relation \eqref{small-comm} is that by choosing
$\h A = \wh{e^{-i 2 \la b}}$ and $\h B = \h \nu$ ---two operators that play
an important role in the Hamiltonian--- we obtain the condition
\be 4 \la Z^3 \wh{e^{-i 2 \la b}}(\v z) \Psi \ll
\wh{e^{-i 2 \la b}}(\v z) \nu(\v z) \Psi, \ee
which becomes
\be \label{large-nu} 4 \la Z^3 \Psi \ll \nu(\v z) \Psi. \ee
when it is multiplied by the operator $\wh{e^{i 2 \la b}}(\v z)$.
This shows that the wave function must have support mostly on large $\nu$.

The condition \eqref{large-nu} has a clear physical meaning: it means that lattice
LQC can only be used so long as the physical volume of each cell remains much
larger than the Planck volume [recall the definition of the volume operator for
each cell given in Eq.\ \eqref{vol-op}].  As the wavelength of the shortest inhomogeneous
mode that can be studied in a given lattice is determined by (twice) the average
physical length of each cell, it follows that, \emph{in lattice LQC, the wavelength
of each mode studied in lattice LQC must remain much larger than $\lp$ at all times.}
This is a strong drawback as it indicates that, for example, lattice LQC cannot be
used if there is inflation.  Nonetheless, lattice LQC can be used in many other
cosmological settings, including the ekpyrotic scenario or matter bounce models.

Another important consequence of the wavelength of all modes remaining much larger
than the Planck scale is that in lattice LQC inverse triad corrections will be
completely negligible.  In \cite{WilsonEwing:2011es}, it is shown how to incorporate
inverse triad corrections in a lattice LQC setting and the main result is that all
nontrivial corrections depend on the ratio of $\lp$ to the wavelength.  Therefore,
if the wavelength is always much larger than the Planck length (as it must be in
lattice LQC), inverse triad effects are completely negligible.

As an aside, we point out that the condition \eqref{large-nu} can
also be derived in the following manner%
\footnote{There is also another way to understand \eqref{large-nu},
coming from the Hamiltonian \eqref{q-ch} defined in
Sec.\ \ref{ss.qm-ham}, together with some results regarding the
superselection sectors in Sec.\ \ref{ss.qm-dyn}.  Restricting
our attention to positive $\nu$ and taking the superselection
sector $\epsilon = 4$ for simplicity (this argument can easily
be extended to any other superselection sector), the wave
function $\Psi(\nu)$ can only have support on $\nu = 4 m \la Z^3$,
with $m \in \mathbb{N}$.
Now, for perturbations to be small, their amplitude must be
much smaller than the mean value of the background.
Due to the superselection sectors, the smallest amplitude of
perturbations possible in $\nu$ is $4 \la Z^3$.  Therefore, for
the perturbations to be much smaller than the background, the
mean value of $\nu$ must be much larger than $4 \la Z^3$.}.
% end footnote
%
%
In the construction of the nonlocal curvature operator used to define
the Hamiltonian, one takes a holonomy around a square loop of area
$\sim \lp^2$.  In order to ensure the straight edges defining this
holonomy remains within one cell, the condition that the length scale
characterizing the cubical cell be larger than $\lp$ must be satisfied,
which is equivalent to demanding that the volume of the cell
be larger than $\lp^3$, and this is precisely \eqref{large-nu}.
Considering the same argument from the point of view of the continuum
(rather than the discrete theory), condition \eqref{large-nu}
can be understood as being equivalent to the assumption that the
connection is varying slowly enough so that gradients of the
connection are negligible for holonomies of length $\lp$.  This
clearly holds for linearized perturbations whose wavelengths are
larger than $\lp$.  Note however that this may no longer be the
case for short ($\lesssim \lp$) wavelength modes in which case
the gradient of the connection could become important and
a different quantization procedure may be necessary.

\subsection{The Hamiltonian and the Constraint Operators}
\label{ss.qm-ham}

In the classical theory, the Hamiltonian as well as the scalar and diffeomorphism
constraints do not exactly vanish and therefore it is impossible to ask that their
operator analogs annihilate physical states in the quantum theory.  Instead, following
the classical theory we will impose that the Hamiltonian and the constraints must be
zero to leading order in the perturbations.  This can be done by determining the strength
of the perturbations (which must be small) by calculating the norm of a wave function
$\Psi$ (satisfying the conditions in Sec.\ \ref{ss.states}) acted upon by an operator
linear in perturbations%
\footnote{Examples of such an operator could be the third term in the scalar constraint
\eqref{q-scal} (the first two terms contribute to both the zeroth and first orders)
or either of the terms in the diffeomorphism constraint \eqref{q-diff}.}.
% end footnote
Denoting this norm by $A^2 \ll 1$, the condition that a constraint $\h \mC$ annihilates
the state $\Psi$ to linear order in the perturbations is that
\be || \h \mC \Psi ||^2 = O(A^4). \ee
Now that it is clear what is meant by a constraint annihilating a state to leading order
in the perturbations, it is possible to define the relevant operators corresponding to
the Hamiltonian and two constraints, all of which annihilate physical states to
first order in perturbation theory.

In order to define the Hamiltonian operator for lattice LQC, we will follow a simple
prescription: we will quantize the homogeneous terms in the Hamiltonian, given in
\eqref{d-chom} just as is done in homogeneous LQC \cite{Ashtekar:2006wn, Ashtekar:2007em}
and the interaction terms \eqref{d-cint} will be quantized in a simple manner determined
by their form in the unique consistent formulation of effective equations in the
longitudinal gauge \cite{WilsonEwing:2011es, Cailleteau:2011kr}.  We will start with
$\mCHhom$, which is easy to quantize as it is basically the Hamiltonian in flat FLRW
models, although the lapse, $N = 4/3 - \nu / 3 \b\nu$, has an unusual form.

In the loop quantization, there are two nontrivial steps that are necessary in order to
obtain the Hamiltonian operator in a homogeneous and isotropic space (which each cell
in the lattice is).  First, as the basic operators in LQC correspond to holonomies
of the connection, it is necessary to express the field strength of the Ashtekar connection.
This is possible as the field strength can be expressed in terms of the holonomy of the connection
around a small square loop.  Due to the discreteness of the area eigenvalue in loop quantum
gravity, it is posited that it is appropriate to choose the area of the square loop to be the
smallest area eigenvalue in loop quantum gravity, denoted by $\la^2 = 4 \sqrt{3} \pi \ga \lp^2$.
A careful implementation of this procedure \cite{Ashtekar:2006wn, Ashtekar:2009vc} shows that
we should express the field strength as, e.g.\
\be \wh {F_{xy}{}^3} = \f{(2 \pi \ga \lp^2)^{2/3}}{\la^2} \, \wh{(\nu^{2/3} \sin^2 \la b)}, \ee
where we have not yet specified a factor-ordering on the right hand side.

The second nontrivial step is to define an inverse triad operator.  This is necessary as the
state $| \nu = 0 \ket$ is normalizable and therefore zero is in the discrete spectrum of the
$\h\nu$ operator.  This difficulty can be avoided in loop quantum gravity by using one of the
Thiemann identities introduced in \cite{Thiemann:1996aw} and a similar approach can be used in LQC.
Although there is an ambiguity in LQC as there are many ways to construct inverse triad
operators (a natural choice for inverse triad operators in lattice LQC models is proposed in
\cite{WilsonEwing:2011es}), all of the inverse triad operators considered so far in the LQC
literature share two properties: first, they annihilate the state $| \nu = 0 \ket$ and
second, their eigenvalues are very close approximations to $1/\nu$ for states $|\nu\ket$
where $V \gg \lp^3$, or, equivalently in lattice LQC, $\nu \gg \lp^3 Z^3$.  Of course,
there can be important inverse triad effects for small, nonzero $\nu$ which may
potentially have observational consequences.  However, as in lattice LQC we necessarily have
$\nu \gg \lp Z^3$ [see \eqref{large-nu}], inverse triad effects are negligible and therefore
we will define the simplest possible inverse triad operator that satisfies the two properties
above:
\be \label{inv-op} \wh{\, \f{1}{|\nu|}\,} |\nu\ket = \begin{cases}
0  & {\rm if} \: \nu = 0, \\
\f{1}{|\nu|} |\nu\ket \quad & {\rm otherwise}. \end{cases} \ee
All other inverse triad operators of different powers can be obtained by raising the operator
defined above to the appropriate power.  Similarly, it is possible to define an inverse triad
operator for the mean operator $\h{\b\nu}$,
\be \wh{\, \f{1}{|\b\nu|}\,} \Psi = \begin{cases}
0  & {\rm if} \: |\b\nu| \, \Psi = 0, \\
\f{1}{|\b\nu|} \Psi \quad & {\rm otherwise}, \end{cases} \ee
where the first case occurs only if $\Psi$ has no support on $\nu_i \neq 0$.  Note that it
is important to take the absolute value as $\Psi$ is symmetric in $\nu$ as can be seen from
Eq.\ \eqref{tot-parity}.

This choice for the inverse triad operator is a simple one, and while it amounts to ignoring
potential effects coming from a ``true'' inverse triad operator, it turns out that the
effect of any other inverse triad operator is extremely well approximated by \eqref{inv-op}
as lattice LQC only provides a valid description of linear cosmological perturbations if
the wavelengths of these perturbations remains considerably larger than the Planck length
in which case the observational consequences of inverse volume operators are completely
negligible as argued above.

Now that the field strength has been expressed in terms of holonomies of the connection
and the inverse triad operators defined, it is possible to define the Hamiltonian
\be \label{q-ch} \wh{\mC_H} = \wh{\mCHhom} + \wh{\mCHint}, \ee
in terms of the homogeneous and interaction terms.  Choosing a factor-ordering where the
lapse is placed on the left and expressing the $\sin^2 \la b$ operator in terms of
complex exponentials,
\begin{align} \label{q-chom}
\wh \mCHhom
%& = \, \f{1}{Z^3} \sum_{\v z} \left(\f{4}{3} - \wh{\f{1}{3|\b\nu|}} |\h\nu(\v z)| \right)
%\bigg[ -\f{3\hbar}{4 \ga \la^2} \wh{\sqrt{|\nu(\v z)|}} \wh{\sin^2 \la b(\v z)}
%\wh{\sqrt{|\nu(\v z)|}} \nn \\
%& \qquad -\f{\hbar}{4 \pi \ga G} \wh{ \f{1}{|\nu(\v z)|} } \f{\del^2}{\del \vp(\v z)^2} \bigg] \nn \\
= \, \f{1}{Z^3} \sum_{\v z} \left(\f{4}{3} - \wh{\f{1}{|\b\nu|}} |\h\nu(\v z)| \right)
\bigg[ \f{3\hbar}{16 \ga \la^2} & \wh{\sqrt{|\nu(\v z)|}} \Big( \wh{e^{-i 2\la b}}(\v z)
+ \wh{e^{i 2\la b}}(\v z) - 2 \Big) \wh{\sqrt{|\nu(\v z)|}} \nn \\
& -\f{\hbar}{4 \pi \ga G} \wh{ \f{1}{|\nu(\v z)|} } \f{\del^2}{\del \vp(\v z)^2}
\bigg]. \end{align}

This is a nice factor-ordering choice as it ensures that, under the action of $\wh \mCHhom$,
the zero volume states decouple from the nonzero volume states.  It also does not couple the
$\nu_i > 0$ and $\nu_i < 0$ sectors (for the relevant superselection sector), which is
necessary in order to preserve the condition \eqref{parity-cond} and also allows us to
restrict our attention to one of the sectors, say $\nu_i > 0$, when we are working with
this part of the Hamiltonian operator.  Then, the sectors where some or all of the
$\nu_i < 0$ are given by Eqs.\ \eqref{tot-parity} and \eqref{parity-cond}.  We will return
to these points at the beginning of Sec.\ \ref{ss.qm-dyn}.

Recent studies of holonomy-corrected effective equations for scalar perturbations in
the longitudinal gauge indicate that there should only be one modification to the
inhomogeneous terms \cite{WilsonEwing:2011es, Cailleteau:2011kr}: the second term in
Eq.\ \eqref{d-cint} should be multiplied by $\cos 2 \la b$.  This gives
\begin{align} \label{q-cint}
\wh \mCHint = \f{1}{Z^3} \sum_{\v z} & \f{\left(2 \pi \ga \lp^2 \right)^{1/3}}
{8 \pi G} |\h \nu(\v z)|^{1/6} \bigg[ \f{1}{9} \wh{\f{1}{\nu(\v z)^2}}
\Big( \v \nabla \h \nu(\v z) \Big)_\dd^2 \nn \\ &
+ 4 \pi G \wh{\cos 2 \la b} \Big( \v \nabla \h \vp(\v z) \Big)_\dd^2 \bigg]
|\h \nu(\v z)|^{1/6}, \end{align}
where we have chosen a symmetric factor ordering in the last term.
The difference terms are defined in Sec.\ \ref{ss.latt}; for example
\be \Big( \v \nabla \h \vp(\v z) \Big)_\dd^2 =
\f{Z^2}{9} \sum_{a=1}^3 \Big( \h \vp(\v z + 2 \h z_a) - \h \vp(\v z - \h z_a) \Big)
\Big( \h \vp(\v z + \h z_a) - \h \vp(\v z - 2 \h z_a) \Big).
\ee

By following the same steps as above, the expression for the scalar constraint operator is
given by
\begin{align} \label{q-scal}
\wh \mH (\v z) = &
\f{3\hbar}{16 \ga \la^2} \wh{\sqrt{|\nu(\v z)|}} \Big( \wh{e^{-i 2\la b}}(\v z)
+ \wh{e^{i 2\la b}}(\v z) - 2 \Big) \wh{\sqrt{|\nu(\v z)|}} \nn \\ &
-\f{\hbar}{4 \pi \ga G} \wh{ \f{1}{|\nu(\v z)|} } \f{\del^2}{\del \vp(\v z)^2}
+ \f{\left(2 \pi \ga \lp^2 \right)^{1/3}}{12 \pi G} \wh{\f{1}{|\nu(\v z)|^{2/3}}}
\Delta_\dd |\h\nu(\v z)|, \end{align}
where the factor-ordering is chosen following $\mCHhom$.

The diffeomorphism constraint operator is slightly more difficult to define.  This is
because an effective analysis of the model shows that the diffeomorphism constraint
should not be modified, even when the curvature is of the Planck scale
\cite{WilsonEwing:2011es, Cailleteau:2011kr}.  Nonetheless, it is necessary to
express all occurances of $b$ in \eqref{d-diff} by a complex exponential.  Luckily,
there is an easy way to ensure that the diffeomorphism constraint is not changed in
the effective limit which is by also multiplying the matter sector of the diffeomorphism
constraint by the appropriate complex exponential of $b$, giving
\be \label{q-diff}
\wh {\mH_a}(\v z) = \f{i\hbar \h\nu(\v z)}{4 \la} (\nabla_\dd)_a \wh{e^{-i 2\la b}}(\v z)
+ \wh{e^{-i 2\la b}}(\v z) \h\pi_\vp(\v z) (\nabla_\dd)_a \h\vp(\v z). \ee
The choice of the factor of 2 in the complex exponential is necessary in order to have
an anomaly-free constraint algebra in the quantum theory.  [See Eq.\ \eqref{diffWeaklyZero}
and the calculations leading to it for an \emph{a posteriori} justification of this.]
An important property of the diffeomorphism constraint defined above is that, since it
annihilates all physical wave functions (to leading order in perturbation theory), so
does the operator
\be \label{q-diff-prime}
\wh {\mH_a'}(\v z) := \wh{e^{i 4\la b}}(\v z) \wh {\mH_a}(\v z)
= -\f{i\hbar \h\nu(\v z)}{4 \la} (\nabla_\dd)_a \wh{e^{i 2\la b}}(\v z)
+ \wh{e^{i 2\la b}}(\v z) \h\pi_\vp(\v z) (\nabla_\dd)_a \h\vp(\v z), \ee
where we have used the relation \eqref{q-prod} twice in order to obtain the right hand
side of the equation above.  Therefore, the diffeomorphism constraint operator can be
defined as either Eq.\ \eqref{q-diff} or \eqref{q-diff-prime} as they are equivalent.

Finally, it is necessary to define the master constraint $\wh {\mC_M}$
\cite{Thiemann:2003zv} as the scalar and diffeomorphism constraints, evaluated
at neighbouring lattice sites, do not commute with themselves or each other%
\footnote{Note that there is an important difference between this master constraint
and the one proposed in \cite{Thiemann:2003zv}: the lattice LQC dynamics are not generated
by the master constraint, but by a separate operator, $\wh{\mC_H}$.  Thus, the role
of the master constraint is more restricted here.}.
This can easily be done with the help of the inverse fiducial metric $\oq^{ab}$,
\be 
\wh {\mC_M} = \sum_{\v z} \left( \wh \mH (\v z) \wh \mH (\v z)
+ \oq^{ab} \wh {\mH_a}(\v z) \wh {\mH_b}(\v z) \right).
\ee
Physical states in lattice LQC are those that are annihilated, to linear order
in the perturbations as defined above, by the master constraint.

The Hamiltonian operator and the scalar, diffeomorphism and master constraint operators
are enough to define the theory of lattice loop quantum cosmology.  In the following part
we shall study the salient aspects of the dynamics and show that all of the constraints
weakly commute with the Hamiltonian.  In Sec.\ \ref{s.eff} we will show that the
resulting effective equations ---for holonomy corrections to scalar perturbations
in LQC--- are the same as those obtained in previous works \cite{WilsonEwing:2011es,
Cailleteau:2011kr}.

\subsection{The Dynamics of Lattice LQC}
\label{ss.qm-dyn}

A nice property of the Hamiltonian operator is that it automatically resolves
the singularity.  If one starts with a nonsingular state, i.e., a wave
function that does not have any support on $\nu(\v z) = 0$ for all $\v z$,
then the wave function will continue to be nonsingular.

This can easily be seen as it is only the shifted terms that allow the support
of the wave function to change.  However, each of terms where the argument has
changed to $\nu(\v n) \pm 4 \la Z^3$ has a prefactor of $|\nu(\v n) \pm 4 \la Z^3|$
(to some positive fractional power) which is clearly equal to zero if the new
argument of the wave function is zero.  Because of this, a wave function with
no initial support on singular states necessarily continues to be nonsingular.
It is in this precise sense that the singularity is resolved by the Hamiltonian.

Also, just as in the flat FLRW case, there are also superselection sectors for
each cell: if one starts with a wave function that only has support on a
particular superselection sector, then the wave function will continue to only
have support on this superselection sector throughout its evolution with respect
to $\wh{\mC_H}$.  The superselection sectors for $\nu(\v n)$ are denoted by
$\epsilon$ and are given by $\epsilon + 4m \la Z^3$ and $4 - \epsilon + 4m \la Z^3$
for $\epsilon \in (0, 2 \la Z^3)$ or $\epsilon + 4m \la N^3$ for $\epsilon$ equal
to $2\la Z^3$ or $4\la Z^3$.  In both cases, $m \in \mathbb{Z}$.  The shorter
stepsize in the superselection sector for $\epsilon \in (0, 2 \la Z^3)$ is due
to the reflection conditions at $\nu = 0$ due to the parity symmetry
$\Psi(|\nu|) = \Psi(\nu)$.

The most interesting superselection sector is the $\epsilon=4$ one as this is
the only superselection sector that includes the singular state $\nu = 0$
(although the singular states do decouple in any case due to the form of the
Hamiltonian).  This superselection sector is also particularly interesting as it
separates the $\nu > 0$ and $\nu < 0$ sectors: it is easy to see that there is
no interaction between these sectors for $\epsilon = 4 \la Z^3$.  Since this
is the only superselection sector that contains the singular state, and it has
the nice property that it separates the positive and negative $\nu$ sectors,
we will work with this superselection sector in each cell for the remainder
of the paper.

Since the positive and negative sectors decouple, it is sufficient to only
consider the part of the wave functions where their arguments are positive.
The negative part of the wave function can be obtained by the parity relation
\eqref{tot-parity}.  In addition, assuming that we start with a nonsingular
state, the $|\nu(\v z) = 0 \rangle$ singular states can be removed from the
Hilbert space as they decouple under the dynamics.  An important ramification
of this is that, once the zero volume states have been removed from the
Hilbert space,
\be \wh { \f{1}{\nu(\v z)} } = \Big( \wh{ \nu(\v z) } \Big)^{-1}. \ee
These simplifications will make it much easier to write out the action of
the Hamiltonian.

Let us now define our notation: we shall denote states by
$\Psi := \Psi(\nu_i; \vp_i)$ where the index $i$ is a shorthand for the vector
$\v z$ denoting the cells in the lattice.  For simplicity of notation, the
arguments of the wave function shall be dropped except when one of the arguments
in the wave function has been shifted.  In this case, only the shifted argument
will be displayed, e.g., $\Psi(\nu_i + 4 \la Z^3)$.

Using the notation described above and restricting our attention to the
$\epsilon = 4$ superselection sector and $\nu > 0$, the action of the
Hamiltonian operator is
$\wh{\mC_H} \Psi = \wh{\mCHhom} \Psi + \wh{\mCHint} \Psi$, where
\begin{align} \wh{\mCHhom}\Psi =&
\f{1}{Z^3} \sum_{i=1}^{Z^3}
\left(\f{4}{3} - \f{\nu_i}{\b\nu} \right)
\bigg[ \f{3\hbar}{16 \ga \la^2} \Big(
\sqrt{\nu_i^2 + 4 \la Z^3 \nu_i} \, \Psi (\nu_i + 4 \la Z^3)  - 2 \nu_i \Psi \nn \\
& \qquad + \sqrt{\nu_i^2 - 4 \la Z^3 \nu_i} \, \Psi (\nu_i - 4 \la Z^3) \Big) 
-\f{\hbar}{4 \pi \ga G \nu_i} \f{\del^2}{\del \vp_i^2} \Psi \bigg],
\end{align}
and
\begin{align} \wh{\mCHint}\Psi =&
\f{1}{Z^3} \sum_{i=1}^{Z^3} \Bigg(
\f{\left(2 \pi \ga \lp^2 \right)^{1/3}}{72 \pi G \nu_i^{5/3}}
\Big( \v \nabla \nu_i \Big)_\dd^2 \Psi
+ \f{\left(2 \pi \ga \lp^2 \nu_i \right)^{1/3}\!\!}{4}
\Big( \v \nabla \vp_i \Big)_\dd^2
\bigg[ (\nu_i + 4 \la Z^3)^{1/6} \Psi (\nu_i + 4 \la Z^3) \nn \\ & \qquad
+ (\nu_i - 4 \la Z^3)^{1/6} \Psi (\nu_i - 4 \la Z^3) \bigg] \Bigg),
\end{align}
where the difference operator is defined in \eqref{def-disc-diff} ---recall
that the index $i$ is simply shorthand for $\v z$--- and it is understood that
\be \f{1}{\b\nu} \Psi = \f{Z^3}{\sum_i \nu_i} \Psi. \ee
It is also possible to explicitly write out the actions of the scalar and
diffeomorphism constraints, but that will not be necessary for our purposes
here.

The physical inner product can be obtained by the group-averaging procedure
\cite{Marolf:1995cn, Giulini:1998rk, Marolf:2000iq} (assuming the Hamiltonian
is essentially self-adjoint),
\be \langle \Psi | \Phi \rangle_{\rm phy} = \f{1}{2\pi} \int_{-\infty}^\infty
\dd \alpha \langle \Psi | e^{i \alpha \wh{\mC_H}} | \Phi \rangle, \ee
while a family of Dirac observables is
given by relational observables with respect to the massless scalar field
$\vp(\v z)$ \cite{Rovelli:2001bz, Dittrich:2004cb, Dittrich:2005kc}.

Alternatively, it is possible to deparametrize the system with respect to
the massless scalar field following \cite{Domagala:2010bm} by using the
diffeomorphism constraint in order to write the Hamiltonian solely in terms
of geometric operators and $\h \pi_\vp(\v z)$.  Then, solving
for $\h \pi_\vp(\v z)$ allows us to use the scalar field $\vp(\v z)$
as a time variable for each cell in the lattice.  This ``emergent'' clock
automatically provides a physical inner product and an explicit definition
of the relational Dirac observables.  We leave this deparametrization as an
exercise for the reader.

Now that the action of the Hamiltonian operator has been written out explicitly,
it is perhaps easier to see some of its properties that were described above.
In particular, it is now more obvious that the singular states $\nu_i = 0$ decouple
from the nonsingular ones under the action of $\wh{\mC_H}$, and also that the
positive and negative $\nu_i$ sectors do not interact.  Finally, the presence
of the superselection sector $\nu = 4 m \la Z^3$, with $m \in \mathbb{N}$ (for
$\nu > 0$) is clear.

Finally, in order for the evolution generated by the Hamiltonian operator to be
compatible with the scalar, diffeomorphism and master constraints, it is necessary
to check that the dynamics are anomaly-free, i.e., that the Poisson bracket of
the constraints with the Hamiltonian are weakly zero to leading order in
perturbation theory.

In order to calculate the relevant commutators, it will be necessary to use several of
the relations obtained in Sec.\ \ref{ss.states} in order to show that the Hamiltonian and
the constraints commute weakly.  In particular, the relations \eqref{q-bar}, \eqref{q-prod},
\eqref{q-poly} and \eqref{large-nu} will be used repeatedly.  We will also only consider
states that have no support on $\nu = 0$ as these states can be removed from the Hilbert
space due to the form of the Hamiltonian.  (Recall that in this case the inverse triad
operator corresponding to $1/\nu$ is the inverse of $\h\nu$.)  Finally, we will drop the
hats on the operators in order to simplify the notation.

Because of its nonlocal nature, commutators where $\h{\b\nu}$ appears require some care
and therefore we will derive the following commutator ---which as we shall see will be a
contribution coming from the lapse--- in some detail.  The relevant nonlocal commutator
to calculate is
\begin{align} \left[ e^{\mp i 2 \la b}(\v z),
-\sum_y \f{\nu(\v y)}{3 \b\nu} \right]
%& = - \sum_y \left( e^{\mp i 2 \la b}(\v z) \f{\nu(\v y)}{3 \b\nu}
%- \f{\nu(\v y)}{3 \b\nu} e^{\mp i 2 \la b}(\v z) \right) \nn \\ &
= \, & - \left( \f{\nu(\v z) \pm 4 \la Z^3}{3 (\b\nu \pm 4 \la)}
- \f{\nu(\v z)}{3 \b\nu} \right) e^{\mp i 2 \la b}(\v z) \nn \\ & \quad
- \sum_{y \neq z} \left( \f{\nu(\v y)}{3 (\b\nu \pm 4 \la)}
- \f{\nu(\v y)}{3 \b\nu} \right) e^{\mp i 2 \la b}(\v z); \nn %\\ &
\end{align}
and it is important to be careful as there is one term in the sum whose commutator is
different from the others.  Then the top term on the right hand side can be separated
into a part which combines with the nonlocal term and a remainder, this gives
\begin{align} \label{psi-comm} \left[ e^{\mp i 2 \la b}(\v z),
-\sum_y \f{\nu(\v y)}{3 \b\nu} \right]
& = \mp \f{4 \la Z^3}{3(\b\nu \pm 4\la)} e^{\mp i 2 \la b}(\v z)
- \sum_y \f{\nu(\v y) [\b\nu - \b\nu \mp 4 \la]}{3 \b\nu (\b\nu \pm 4 \la)}
e^{\mp i 2 \la b}(\v z)  \nn \\ &
= \mp \f{4 \la Z^3}{3} e^{\mp i 2 \la b}(\v z) \f{1}{\b\nu}
\pm \f{4 \la}{3 \b\nu} \sum_y \nu(\v y) e^{\mp i 2 \la b}(\v z) \f{1}{\b\nu}.
\end{align}
An important observation here is that there is a nonlocal term on the right hand side due
to the nontrivial commutator of $\h{\b\nu}^{-1}$ and the shift operator which could be
problematic.  However, as we shall see shortly, although such a nonlocal contribution does
indeed appear in the commutators between the Hamiltonian and the constraints, it will be
multiplied by the scalar constraint (modulo a term which sums to zero) and therefore does
not pose any problems.

The commutator between the scalar constraint and the Hamiltonian is given by
\begin{align} \left[ \mH(\v z), \mC_H \right] =&
\f{1}{Z^3} \sum_y \left[ \mH^1(\v z), -\f{\nu(\v y)}{3 \b\nu} \right]
\Big( \mH^1(\v y) + \mH^2(\v y) \Big) \nn \\ &
+ \f{1}{Z^3} \sum_y \left( \f{4}{3} - \f{\nu(\v y)}{3 \b\nu} \right)
\bigg( \left[ \mH^1(\v z), \mH^2(\v y) \right]
+ \left[ \mH^2(\v z), \mH^1(\v y) \right]
+ \left[ \mH^3(\v z), \mH^1(\v y) \right] \bigg) \nn \\ &
+ \left[ \mH^1(\v z), {\mCHint}^{(1)} \right]
+ \left[ \mH^2(\v z), {\mCHint}^{(2)} \right],
\end{align}
where the superscripts indicate the first, second and third terms of each operator
as defined in \eqref{q-chom}, \eqref{q-cint} and \eqref{q-scal}.  There are
several terms that obviously commute (either exactly or to first order in the
perturbations) and these terms have been dropped.  In addition, it is easy to
see that the first two terms on the second line cancel.  We shall now calculate
the remaining terms one by one, to first order in perturbation theory as all
higher order terms can be dropped.

The first term can easily be computed by using the result \eqref{psi-comm},
and gives
\begin{align} \f{1}{Z^3} \sum_y \left[ \mH^1(\v z), -\f{\nu(\v y)}{3 \b\nu} \right] &
\Big( \mH^1(\v y) + \mH^2(\v y) \Big) = \,
\f{i \hbar \sqrt{\nu(\v z)}}{2 \ga \la} \sin 2 \la b(\v z) \f{\sqrt{\nu(\v z)}}{\b\nu}
\Big( \mH^1(\v z) + \mH^2(\v z) \Big) \nn \\ &
- \f{i \hbar \sqrt{\nu(\v z)}}{2 \ga \la Z^3 \b\nu} \sum_y \nu(\v y)
\sin 2 \la b(\v z) \f{\sqrt{\nu(\v z)}}{\b\nu}
\Big( \mH^1(\v y) + \mH^2(\v y) \Big).
\end{align}
The second line of this result can be simplified by noting that
\be \sum_y \nu(\v y) \sin 2 \la b(\v z) \f{\sqrt{\nu(\v z)}}{\b\nu}
\mH^3(\v y) = \nu(\v z) \sin 2 \la b(\v z) \f{\sqrt{\nu(\v z)}}{\b\nu}
\sum_y \mH^3(\v y) = 0. \ee
The first equality holds to first order in the perturbations as there is a difference
operator in $\mH^3$, while the second equality holds as the sum over all cells of a
difference operator vanishes (this is the discrete equivalent of Stokes' theorem on
a manifold without boundaries).  Therefore, by adding this particular zero to the
commutator, we find that
\begin{align} \label{comm-s1} \f{1}{Z^3} \sum_y \left[ \mH^1(\v z), -\f{\nu(\v y)}{3 \b\nu} \right]
\Big( \mH^1(\v y) + \mH^2(\v y& ) \Big) =
\f{i \hbar \sqrt{\nu(\v z)}}{2 \ga \la} \sin 2 \la b(\v z) \f{\sqrt{\nu(\v z)}}{\b\nu}
\Big( \mH^1(\v z) + \mH^2(\v z) \Big) \nn \\ &
- \f{i \hbar \sqrt{\nu(\v z)}}{2 \ga \la Z^3 \b\nu} \sum_y \nu(\v y)
\sin 2 \la b(\v z) \f{\sqrt{\nu(\v z)}}{\b\nu} \mH(\v y),
\end{align}
where it is now clear that the nonlocal term is weakly zero.

The second nontrivial commutator,
\begin{align} \f{1}{Z^3} \sum_y \left( \f{4}{3} - \f{\nu(\v y)}{3 \b\nu} \right)
\left[ \mH^3(\v z), \mH^1(\v y) \right] &=
\f{(2 \pi \ga\lp^2)^{1/3}}{12 \pi G Z^3} \sum_y \left( \f{4}{3} - \f{\nu(\v y)}{3 \b\nu} \right)
\left[ \f{1}{\nu(\v z)^{2/3}} \Delta_\dd \nu(\v z), \mH^1(\v y) \right] \nn \\ &
= \f{(2 \pi \ga\lp^2)^{1/3}}{12 \pi G Z^3 \nu(\v z)^{2/3}}
\sum_y \left( \f{4}{3} - \f{\nu(\v y)}{3 \b\nu} \right)
\left[ \Delta_\dd \nu(\v z), \mH^1(\v y) \right] \nn \\ &
\quad + \f{(2 \pi \ga\lp^2)^{1/3}}{12 \pi G Z^3} \sum_y
\left[ \f{1}{\nu(\v z)^{2/3}}, \mH^1(\v y) \right] \Delta_\dd \nu(\v z),
\end{align}
splits naturally into two separate terms.  The first term will pick out 6
contributions from the sum due to the structure of the difference operator
\eqref{def-disc-lapl} (that can themselves be expressed in terms of $\Delta_\dd$).
One might be tempted to drop the second term as it is of the form \eqref{drop-comm},
but as the prefactor is very large [see \eqref{large-nu} and remember the form of
$\mH^1$], this term is relevant.  A straightforward calculation shows that the
first term gives
\begin{align}
%\f{(2 \pi \ga\lp^2)^{1/3}}{12 \pi G Z^3 \nu(\v z)^{2/3}}
%\sum_y \left( \tf{4}{3} - \tf{\nu(\v y)}{3 \b\nu} \right)
%\left[ \Delta_\dd \nu(\v z), \mH^1(\v y) \right] =&
\f{i \hbar (2 \pi \ga \lp^2)^{1/3}}{8 \pi G \ga \la \nu(\v z)^{2/3}}
\Delta_\dd \left( \f{4}{3} - \f{\nu(\v z)}{3 \b\nu} \right)
\sqrt{\nu(\v z)} \sin 2 \la b & (\v z) \sqrt{\nu(\v z)} = %\nn \\ =&
\f{i \hbar (2 \pi \ga \lp^2)^{1/3} \nu(\v z)^{1/3}}{8 \pi G \ga \la}
\Delta_\dd \sin 2 \la b (\v z) \nn \\ &
+ \f{i \hbar (2 \pi \ga \lp^2)^{1/3}}{12 \pi G \ga \la \nu(\v z)^{2/3}}
\sin 2 \la b (\v z) \Delta_\dd \nu(\v z),
\label{comm31-1}
\end{align}
which has been rewritten on the right hand side by using the relations \eqref{q-bar},
\eqref{q-prod} and \eqref{q-poly}.

The second term can also be calculated rather easily.  It is easy to see that
the contribution from the second term is given by
\begin{align}
%\f{(2 \pi \ga\lp^2)^{1/3}}{12 \pi G Z^3} \sum_y
%\left[ \f{1}{\nu(\v z)^{2/3}}, \mH^1(\v y) \right] \Delta_\dd \nu(\v z) =&
\f{(2 \pi \ga\lp^2)^{1/3}}{12 \pi G Z^3} & \left( \f{1}{\nu(\v z)^{2/3}}
- \f{1}{[\nu(\v z) + 4 \la Z^3]^{2/3}} \right) e^{-i 2 \la b} (\v z)
\Delta_\dd \nu(\v z) \nn \\ &
+ \f{(2 \pi \ga\lp^2)^{1/3}}{12 \pi G Z^3} \left( \f{1}{\nu(\v z)^{2/3}}
- \f{1}{[\nu(\v z) - 4 \la Z^3]^{2/3}} \right) e^{i 2 \la b} (\v z)
\Delta_\dd \nu(\v z),
\end{align}
which can be simplified by Taylor-expanding $(\nu \pm 4 \la Z^3)^{-2/3}$
to second order [all of the higher order terms can be neglected due
to \eqref{large-nu}].  This gives
\be -\f{i \hbar (2 \pi \ga \lp^2)^{1/3}}{12 \pi G \ga \la \nu(\v z)^{2/3}}
\sin 2 \la b (\v z) \Delta_\dd \nu(\v z), \label{comm31-2} \ee
which will cancel part of the contribution from the first term.

Combining the results \eqref{comm31-1} and\eqref{comm31-2}, we find that
\be \label{comm-s2} \f{1}{Z^3} \sum_y \left( \f{4}{3} - \f{\nu(\v y)}{3 \b\nu} \right)
\left[ \mH^3(\v z), \mH^1(\v y) \right] =
\f{i \hbar (2 \pi \ga \lp^2)^{1/3} \nu(\v z)^{1/3}}{8 \pi G \ga \la}
\Delta_\dd \sin 2 \la b (\v z). \ee

There are two commutators that remain, both involving the interaction
terms in the Hamiltonian.  These are both relatively easy to calculate.
By using the relations \eqref{q-bar}, \eqref{q-prod} and \eqref{q-poly},
it is possible to express these commutators as
\be \label{comm-s3} \left[ \mH^1(\v z), {\mCHint}^{(1)} \right] =
\f{i \hbar \sqrt{\nu(\v z)}}{2 \ga \la} \sin 2 \la b(\v z) \f{\sqrt{\nu(\v z)}}{\b\nu}
\mH^3(\v z), \ee
\be \label{comm-s4} \left[ \mH^2(\v z), {\mCHint}^{(2)} \right] =
\f{i \hbar}{(2 \pi \ga \lp^2)^{2/3} \nu(\v z)^{2/3}} \cos 2 \la b(\v z)
\pi_\vp(\v z) \Delta_\dd \vp(\v z).
\ee
The awkward factor-ordering in \eqref{comm-s3} is chosen to coincide with
that in \eqref{comm-s1}.

By combining Eqs.\ \eqref{comm-s1}, \eqref{comm-s2}, \eqref{comm-s3} and
\eqref{comm-s4}, we find that
\begin{align} \left[ \mH(\v z), \mC_H \right] =&
\f{i \hbar \sqrt{\nu(\v z)}}{2 \ga \la} \sin 2 \la b(\v z) \f{\sqrt{\nu(\v z)}}{\b\nu}
\mH(\v z) 
+ \f{i \hbar}{2 (2 \pi \ga \lp^2)^{2/3} \nu(\v z)^{2/3}} \Big[ \mH_a(\v z)
+ \mH_a'(\v z) \Big] \nn \\ &
- \f{i \hbar \sqrt{\nu(\v z)}}{2 \ga \la Z^3 \b\nu} \sum_y \nu(\v y)
\sin 2 \la b(\v z) \f{\sqrt{\nu(\v z)}}{\b\nu} \mH(\v y),% \nn \\ &
\end{align}
thus showing that the scalar constraint weakly commutes with the Hamiltonian
to first order in perturbation theory.

The next step is to calculate the commutator between the diffeomorphism constraint
$\mH_a$ and the Hamiltonian [it is easy to see by the first equality in
\eqref{q-diff-prime} that if $\mH_a$ commutes weakly with the Hamiltonian, so
will $\mH_a'$]:
\begin{align} \label{comm-diff1}
\left[ \mH_a(\v z), \mC_H \right] =&
\f{i \hbar}{4 \la} \left[ \nu(\v z), \mC_H \right]
(\nabla_\dd)_a e^{-i 2 \la b}(\v z)
+ \f{i \hbar}{4 \la} \nu(\v z)
\left[ (\nabla_\dd)_a e^{-i 2 \la b}(\v z), \mC_H \right] \nn \\ &
+ \pi_\vp(\v z) \Big( (\nabla_\dd)_a \vp(\v z) \Big)
\left[ e^{-i 2 \la b}(\v z), \mC_H \right]
+ \pi_\vp(\v z) \left[ (\nabla_\dd)_a \vp(\v z), \mC_H \right]
e^{-i 2 \la b}(\v z) \nn \\ &
+ \left[ \pi_\vp(\v z), \mC_H \right] \Big( (\nabla_\dd)_a \vp(\v z) \Big)
e^{-i 2 \la b}(\v z).
\end{align}
In order to complete this calculation, we shall compute one by one the commutators
of $\vp(\v z), \pi_\vp(\v z), \nu(\v z)$ and $\exp(-i 2 \la b)(\v z)$ with the
Hamiltonian.  The second and third of these commutators need only be calculated
to leading order due to the presence of a difference operator multiplying each
of the appearances of these commutators, while the other two commutators must
be calculated to first order in perturbations.

The two commutators that are only needed to leading order are both
easily calculated,
\begin{align}
\label{comm-nu}
\left[ \nu(\v z), \mC_H \right] &= -\f{3 \hbar}{4 \ga \la}
\sqrt{\nu(\v z)} \Big( e^{-i 2 \la b}(\v z) - e^{i 2 \la b}(\v z) \Big)
\sqrt{\nu(\v z)}, \\
\label{comm-pi}
\left[ \pi_\vp(\v z), \mC_H \right] &= 0.
\end{align}
Of the two remaining commutators, both of which must be calculated to first
order in perturbations, $[\vp, \mC_H]$ is the easier calculation and gives
\be \label{comm-phi}
\left[ \vp(\v z), \mC_H \right] = \left(\f{4}{3} - \f{\nu(\v z)}{3 \b\nu}\right)
\f{i \hbar \pi_\vp(\v z)}{2 \pi \ga \lp^2 \nu(\v z)}.
\ee
The last commutator,
\begin{align}
\left[ e^{-i 2 \la b}(\v z), \mC_H \right] =&
\f{1}{Z^3} \sum_y \left( \f{4}{3} -\f{\nu(\v y)}{3 \b\nu} \right)
\left[ e^{-i 2 \la b}(\v z), \mH^1(\v y) + \mH^2(\v y) \right]
+ \left[ e^{-i 2 \la b}(\v z), {\mCHint}^{(1)} \right]  \nn \\ &
+ \f{1}{Z^3} \sum_y \left[ e^{-i 2 \la b}(\v z), -\f{\nu(\v y)}{3 \b\nu} \right]
\Big( \mH^1(\v y) + \mH^2(\v y) \Big),
\end{align}
is a little more complicated and shall be calculated term by term.
Note that although ${\mCHint}^{(2)}$ does not exactly commute with the
shift operator, it does up to first order in the perturbations.

The first term $(1/Z^3) \sum_y ( 1 + \psi(\v y) )
\left[ e^{-i 2 \la b}(\v z), \mH^1(\v y) + \mH^2(\v y) \right]$ gives
\begin{align}
%\f{1}{Z^3} \sum_y & \left( \f{4}{3} -\f{\nu(\v y)}{3 \b\nu} \right)
%\left[ e^{-i 2 \la b}(\v z), \mH^1(\v y) + \mH^2(\v y) \right] = \nn \\ =
\bigg( & \f{4}{3} - \f{\nu(\v z)}{3 \b\nu} \bigg) \Bigg\{ \f{3 \hbar}{16 \ga \la^2 Z^3}
\bigg( \sqrt{\nu(\v z) + 4 \la Z^3} \big( e^{-i 2 \la b}(\v z) + e^{i 2 \la b}(\v z)
- 2 \big) \sqrt{\nu(\v z) + 4 \la Z^3} \nn \\ &
- \sqrt{\nu(\v z)} \big( e^{-i 2 \la b}(\v z) + e^{i 2 \la b}(\v z) - 2 \big)
\sqrt{\nu(\v z)} \bigg)
- \f{\la \pi_\vp(\v z)^2}{\pi \ga \lp^2 \nu(\v z) [\nu(\v z) + 4 \la Z^3]} \Bigg\}
e^{-i 2 \la b}(\v z),
\end{align}
while the contribution from the second commutator is
\begin{align} \label{comm-shift2}
\left[ e^{-i 2 \la b}(\v z), {\mCHint}^{(1)} \right] = - \f{4 \la}{3 \nu(\v z)}
\mH^3(\v z) e^{-i 2 \la b}(\v z).
\end{align}
Note that the factor-ordering can be changed freely due to relation \eqref{drop-comm}
as there is a difference operator in $\mH^3$.

Finally, the last commutator gives
\begin{align} \label{comm-shift3}
\f{1}{Z^3} \sum_y \left[ e^{-i 2 \la b}(\v z), -\f{\nu(\v y)}{3 \b\nu} \right]
\Big( \mH^1(\v y) + \mH^2(\v y) \Big) = &
\f{4 \la}{3 Z^3 \b\nu} \sum_y \nu(\v y) e^{-i 2 \la b}(\v z) \f{1}{\b\nu}
\Big( \mH^1(\v y) + \mH^2(\v y) \Big) \nn \\ &
- \f{4 \la}{3} e^{-i 2 \la b}(\v z) \f{1}{\b\nu} \Big( \mH^1(\v z) + \mH^2(\v z) \Big),
\end{align}
and it is possible to combine all of these terms which together show that
\begin{align} \label{comm-shift}
\Big[ e & {}^{-i 2 \la b}(\v z), \mC_H \Big] =
\f{4 \la}{3 Z^3 \b\nu} \sum_y \nu(\v y) e^{-i 2 \la b}(\v z) \f{1}{\b\nu} \mH(\v y)
- \f{4 \la}{3} e^{-i 2 \la b}(\v z) \f{1}{\b\nu} \mH(\v z) \nn \\ &
+ \bigg( \f{4}{3} - \f{\nu(\v z)}{3 \b\nu} \bigg) \Bigg\{ \f{3 \hbar}{16 \ga \la^2 Z^3}
\bigg( \sqrt{\nu(\v z) + 4 \la Z^3} \big( e^{-i 2 \la b}(\v z) + e^{i 2 \la b}(\v z)
- 2 \big) \sqrt{\nu(\v z) + 4 \la Z^3} \nn \\ &
- \sqrt{\nu(\v z)} \big( e^{-i 2 \la b}(\v z) + e^{i 2 \la b}(\v z) - 2 \big)
\sqrt{\nu(\v z)} \bigg) e^{-i 2 \la b}(\v z)
- \f{\la \pi_\vp(\v z)^2}{\pi \ga \lp^2 \nu(\v z)} e^{-i 2 \la b}(\v z)
\f{1}{\nu(\v z)}
\Bigg\},
\end{align}
where we have changed the factor-ordering in \eqref{comm-shift2} by using the
relation \eqref{drop-comm} and we have added zero to the nonlocal term in
\eqref{comm-shift3} in the form of
\be \f{4 \la}{3 Z^3 \b\nu} \sum_y \nu(\v y) e^{-i 2 \la b}(\v z)
\f{1}{\b\nu} \mH^3(\v y) =
\f{4 \la}{3 Z^3 \b\nu} \nu(\v z) e^{-i 2 \la b}(\v z)
\f{1}{\b\nu} \sum_y \mH^3(\v y) = 0, \ee
in order to show explicitly that the nonlocal term is zero on the constraint surface.
This expression vanishes as the sum of a difference operator over the entire lattice
is identically zero.

It is now possible to complete the calculation of the commutator, started in Eq.\
\eqref{comm-diff1}, between the diffeomorphism constraint and the Hamiltonian by using
the results \eqref{comm-nu}, \eqref{comm-pi}, \eqref{comm-phi} and \eqref{comm-shift}.
We will not go through the calculation step by step as it is not particularly difficult
(although a little tedious), but rather present two relations that are necessary for
this derivation that may not be immediately obvious.  The first, which can be derived
from the relation \eqref{q-prod}, is
\begin{align}
\sqrt{\nu(\v z) + 4 \la Z^3} (\nabla_\dd)_a \sqrt{\nu(\v z) + 4 \la Z^3}
= & \, \f{1}{2} (\nabla_\dd)_a \Big( \nu(\v z) + 4 \la Z^3 \Big) \nn \\ = & \,
\f{1}{2} (\nabla_\dd)_a \nu(\v z) \nn \\ = & \,
\sqrt{\nu(\v z)} (\nabla_\dd)_a \sqrt{\nu(\v z)},
\end{align}
and therefore the difference between the term on the left and the bottom term on the
right is zero.  Two other terms that can also be seen to be cancel by using relation
\eqref{q-prod} are
\be e^{-i 2 \la b}(\v z) (\nabla_\dd)_a e^{i 2 \la b}(\v z)
+ e^{i 2 \la b}(\v z) (\nabla_\dd)_a e^{-i 2 \la b}(\v z) = 
(\nabla_\dd)_a \Big( e^{i 2 \la b}(\v z) e^{-i 2 \la b}(\v z) \Big) = 0. \ee
After a long but straightforward calculation where the relations \eqref{q-bar},
\eqref{q-prod}, \eqref{q-poly} and \eqref{drop-comm} are used repeatedly, as
well as the two equalities derived above, the result
\begin{align} \label{diffWeaklyZero}
\left[ \mH_a(\v z), \mC_H \right] =& 
- \f{i \hbar}{3} (\nabla_\dd)_a
\left[ e^{-i 2 \la b}(\v z) \mH(\v z)
- \f{1}{Z^3} \sum_y \nu(\v y) e^{-i 2 \la b}(\v z)
\f{1}{\b\nu} \mH(\v y) \right] \nn \\ &
+ \left[ \f{3 \hbar}{4 \ga \la} \left( e^{-i 2 \la b}(\v z)
+ e^{i 2 \la b}(\v z) - 2 \right)
- \f{\la \pi_\vp(\v z)^2}{\pi \ga \lp^2 \nu(\v z)^2}
\right] \mH_a(\v z) \nn \\ &
- \f{i \hbar}{3} e^{-i 2 \la b}(\v z)
\f{ (\nabla_\dd)_a \nu(\v z)}{\nu(\v z)} \mH(\v z),
\end{align}
is obtained, showing that the diffeomorphism constraint weakly commutes with the
Hamiltonian to leading order in perturbation theory.

These calculations show that in lattice LQC the quantum dynamics generated by the
Hamiltonian preserve the scalar and diffeomorphism constraints, and therefore the
master constraint also weakly commutes with the Hamiltonian.  This shows that
lattice LQC is anomaly-free.  

There are two main limitations here.  The first is that we are working in a
perturbative setting and therefore we can only show that lattice LQC is
anomaly-free to first order in perturbations.  Nonetheless, this is the
best that can be done when working with linear perturbations, as we have been
from the very beginning.  The other limitation is that the physical volume of
each cell in the lattice must remain much larger than the Planck volume at
all times [see condition \eqref{large-nu}].  The main consequence of this is
that short-wavelength modes (with respect to $\lp$) cannot be modelled in lattice
LQC, and therefore inverse triad corrections cannot be studied here.  One would
expect contributions from inverse triad corrections to play a potentially
important role for short wavelengths, but to understand these it will be necessary
to use a different approach to study linear perturbations in LQC than the lattice
model presented in this paper.

\section{Effective Equations}
\label{s.eff}

The effective theory is obtained by studying the dynamics of states that are
sharply peaked.  The resulting effective equations should provide the leading
order quantum corrections to the classical dynamics.  In general, we do not
expect the effective equations to be valid when quantum effects are at their
most important, but surprisingly the effective equations provide excellent
approximations to the full quantum dynamics of sharply peaked states ---even
when quantum effects are strongest--- for FLRW models.

As in lattice LQC each cell is homogeneous and isotropic, and the interactions
between nearby cells are small, it is reasonable to hope that the effective
theory will be just as good an approximation to the fully quantum dynamics of
sharply peaked wave functions as in the exactly homogeneous case.

We will first obtain the effective equations for lattice LQC and then the
continuum limit will be very easy to take, giving the holonomy-corrected
effective equations for scalar perturbations in LQC.  (Recall that inverse triad
corrections are not included due to the condition \eqref{large-nu} which restricts
the wavelength of the modes considered in the lattice to always remain much
greater than the Planck length, in which case inverse triad corrections are
expected to be negligible.)

\subsection{The Effective Theory on the Lattice}
\label{ss.eff-latt}

The Hamiltonian for the effective equations is given by the sum of the terms
corresponding to the homogeneous and interaction parts of the Hamiltonian
given in \eqref{q-chom} and \eqref{q-cint},
\begin{align} \label{e-ham}
\mC_H = \, & \f{1}{Z^3} \sum_{\v z} \left(\f{4}{3} - \f{\nu(\v z)}{3\b\nu} \right)
\bigg[ - \f{3\hbar}{4 \ga \la^2} \nu(\v z) \sin^2 \la b(\v z)
+ \f{\pi_\vp^2(\v z)}{4 \pi \ga \lp^2 \nu(\v z)} \bigg] \nn \\ &
+ \f{1}{Z^3} \sum_{\v z} \f{\left[2 \pi \ga \lp^2 \nu(\v z) \right]^{1/3}}
{8 \pi G} \bigg[ \f{1}{9 \nu(\v z)^2} \Big( \v \nabla \nu(\v z) \Big)_\dd^2
+ 4 \pi G \cos 2 \la b(\v z) \Big( \v \nabla \vp(\v z) \Big)_\dd^2 \bigg], \end{align}
except that $\mC_H$ (and its components) are now treated as classical objects.
The scalar and diffeomorphism constraints are similarly the classical equivalents
to the operator equations \eqref{q-scal} and \eqref{q-diff},
\be \label{e-scal}
\mH (\v z) = - \f{3\hbar}{4 \ga \la^2} \nu(\v z) \sin^2 \la b(\v z)
+ \f{\pi_\vp^2(\v z)}{4 \pi \ga \lp^2 \nu(\v z)}
+ \f{\left(2 \pi \ga \lp^2 \right)^{1/3}}{12 \pi G \nu(\v z)^{2/3}}
\Delta_\dd \nu(\v z) \approx 0, \ee
and
\be \label{e-diff}
\mH_a(\v z) = \f{\hbar \nu(\v z)}{2} (\nabla_\dd)_a b(\v z)
+ \pi_\vp(\v z) (\nabla_\dd)_a \vp(\v z) \approx 0. \ee
Note that the $e^{-i 2 \la b}$ terms cancel in $\mH_a$ and therefore the
diffeomorphism constraint in the effective theory is the same as the classical
discretized diffeomorphism constraint \eqref{d-diff}.  On the other hand, the
scalar constraint is different in the effective theory.

The effective dynamics are obtained by following the same procedure as usual
with the Hamiltonian and constraints given above.  Recalling the Poisson brackets
on the lattice \eqref{poisson1} and \eqref{poisson2}, we find that the effective
equations are given by
\begin{align}
\d b(\v z) &= - \Big( 1 + \psi(\v z) \Big) \left( \f{3}{2 \ga \la^2} \sin^2 \la b(\v z)
+ \f{\pi_\vp(\v z)^2}{2 \pi \ga \lp^2 \hbar \nu(\v z)^2} \right), \\
\d \nu(\v z) &= \Big( 1 + \psi(\v z) \Big)
\f{3}{\ga \la} \nu(\v z) \sin \la \b(\v z) \cos \la b(\v z), \\
\d \vp(\v z) &= \Big( 1 + \psi(\v z) \Big)
 \f{\pi_\vp(\v z)}{2 \pi \ga \lp^2 \nu(\v z)}, \\
\d \pi_\vp(\v z) &= \Big( 2 \pi \ga \lp^2 \nu(\v z) \Big)^{1/3}
\cos 2 \la b(\v z) \Delta_\dd \vp(\v z),
\end{align}
where it is understood that
\be \psi(\v z) = \f{4}{3} - \f{\nu(\v z)}{3 \b\nu}. \ee
It is possible to show that these equations of motion conserve the constraints,
i.e., the Hamiltonian weakly commutes with the scalar and diffeomorphism
constraints \cite{WilsonEwing:2011es, Cailleteau:2011kr}.

\subsection{The Continuum Limit}
\label{ss.eff-cont}

The lattice is essential in the quantum theory in order to have available all
of the LQC tools that have been studied in considerable detail for homogeneous
space-times.  Furthermore it is not clear how to obtain, as a limit from
lattice LQC, a quantum theory defined on a continuum, or even if such a
limit exists.  However, the situation is very different in the effective theory
where the continuum limit is trivial.
%
%In fact, the effective dynamics were first studied in the continuum without even
%knowing the quantum theory \cite{WilsonEwing:2011es, Cailleteau:2011kr}.  We will
%see that the continuum limit of the effective theory obtained in
%Sec.\ \ref{ss.eff-latt} is identical to the effective equations
%studied in Refs.\ \cite{WilsonEwing:2011es, Cailleteau:2011kr}.

The effective Hamiltonian, in the continuum, becomes
\begin{align} \label{ec-ham}
\mC_H = \, & \int_\mM \dd^3 \v x \left(\f{4}{3} - \f{\nu(\v x)}{3\b\nu} \right)
\bigg[ - \f{3\hbar}{4 \ga \la^2} \nu(\v x) \sin^2 \la b(\v x)
+ \f{\pi_\vp^2(\v x)}{4 \pi \ga \lp^2 \nu(\v x)} \bigg] \nn \\ &
+ \int_\mM \dd^3 \v x \f{\left[2 \pi \ga \lp^2 \nu(\v x) \right]^{1/3}}
{8 \pi G} \bigg[ \f{1}{9 \nu(\v x)^2} \Big( \v \nabla \nu(\v x) \Big)^2
+ 4 \pi G \cos 2 \la b(\v x) \Big( \v \nabla \vp(\v x) \Big)^2 \bigg]. \end{align}
Similarly, the scalar and diffeomorphism constraints are
\be \label{ec-scal}
\mH (\v x) = - \f{3\hbar}{4 \ga \la^2} \nu(\v x) \sin^2 \la b(\v x)
+ \f{\pi_\vp^2(\v x)}{4 \pi \ga \lp^2 \nu(\v x)}
+ \f{\left(2 \pi \ga \lp^2 \right)^{1/3}}{12 \pi G \nu(\v x)^{2/3}}
\Delta \nu(\v x) \approx 0, \ee
and
\be \label{ec-diff}
\mH_a(\v x) = \f{\hbar \nu(\v x)}{2} \nabla_a b(\v x)
+ \pi_\vp(\v x) \nabla_a \vp(\v x) \approx 0, \ee
in the continuum.  The equations of motion can be obtained from the Hamiltonian
\eqref{ec-ham} and using the Poisson brackets \eqref{poiss1} and \eqref{poiss2},
giving
\begin{align}
\label{e-bdot} \d b(\v x) &= - \Big( 1 + \psi(\v x) \Big) \left(
\f{3}{2 \ga \la^2} \sin^2 \la b(\v x)
+ \f{\pi_\vp(\v x)^2}{2 \pi \ga \lp^2 \hbar \nu(\v x)^2} \right), \\
\label{e-nudot} \d \nu(\v x) &= \Big( 1 + \psi(\v x) \Big) \f{3}{\ga \la} \nu(\v x)
\sin \la b(\v x) \cos \la b(\v x), \\
\label{e-phidot} \d \vp(\v x) &= \Big( 1 + \psi(\v x) \Big)
 \f{\pi_\vp(\v x)}{2 \pi \ga \lp^2 \nu(\v x)}, \\
\label{e-pidot} \d \pi_\vp(\v x) &= \Big( 2 \pi \ga \lp^2 \nu(\v x) \Big)^{1/3}
\cos 2 \la b(\v x) \Delta \vp(\v x).
\end{align}

In the (continuous) effective theory, the scalar and diffeomorphism
constraints \eqref{ec-scal} and \eqref{ec-diff} must be satisfied and
then the evolution of the basic variables generated by the Hamiltonian
\eqref{ec-ham} (which weakly commutes with the constraints) is given by
Eqs.\ \eqref{e-bdot}---\eqref{e-pidot}.  This picture is nicely consistent
with previous studies of the holonomy-corrected effective equations for
linear cosmological perturbations as these equations are in exact agreement
with those obtained in \cite{WilsonEwing:2011es, Cailleteau:2011kr}.

Finally, the correct classical limit is easily obtained from these equations
by sending $\hbar \to 0$ (recall that there is an $\hbar$ in $\la$, and
see \cite{WilsonEwing:2011es} for the classical equations of motion for
the variables $b$ and $\nu$).

\section{Discussion}
\label{s.dis}

Lattice LQC is an anomaly-free loop quantization of linear cosmological
perturbations.  It includes the same quantum geometry effects that cause the
bounce in homogeneous models and it is easy to see that the singular states
decouple from nonsingular ones under the action of the Hamiltonian, thus
resolving the classical singularity.  In addition, the effective dynamics of
lattice LQC are the same as those derived for holonomy corrections in Refs.\
\cite{WilsonEwing:2011es, Cailleteau:2011kr}, and the correct classical
limit is obtained when we take $\hbar \to 0$ in the effective equations.

We expect the effective equations to provide an excellent approximation to
the full dynamics of sharply peaked states, just as in the homogeneous case,
because in lattice LQC each cell is taken to be a flat FLRW model which
interacts only weakly with its neighbours.  Therefore, it seems reasonable
to expect that the effective equations will continue to be very good
approximations to the quantum dynamics of sharply-peaked states.  Nonetheless,
this hypothesis should be tested; this would probably be easiest to do
numerically.  Numerical studies would also be very useful to verify that
the bounce observed in homogeneous models remains also present in lattice
LQC, as well as studying the matter energy density $\rho$.  In homogeneous
models, $\rho$ is bounded above by the critical energy density
$\rho_c = 0.41 \rho_{\rm Pl}$, and it would be interesting to see if the same
holds in lattice LQC, possibly modulo some small fluctations.

The canonical quantization in lattice LQC is more complicated than
that of homogeneous models due to, among other things, the presence of
the diffeomorphism constraint in addition to the scalar constraint.  A
potentially important point is that the scalar and diffeomorphism
constraints are treated quite differently in lattice LQC.  The simplest
way to see this is by looking at the effective theory, where the
diffeomorphism constraint is the same as in the classical theory,
but the scalar constraint is significantly different.  In other words,
the classical diffeomorphism symmetry is not modified even when the
space-time curvature nears the Planck scale.  This might indicate
that time-like and space-like diffeomorphisms are on a different
footing in the quantum theory: the space-like diffeomorphism symmetry
seems to continue to hold in quantum gravity, while the time-like
diffeomorphism symmetry is modified when the curvature becomes large.
It is not clear if this property is simply the artefact of a simple
model or whether it which holds more generally, but it is certainly
worth further thought.

In this paper, we have seen how scalar perturbations can be studied in
lattice LQC.  However, it is not as easy to put vector and tensor
perturbations on a lattice.  This is because the metric of a flat FLRW
space-time with vector or tensor perturbations is necessarily nondiagonal.
Nondiagonal metrics are difficult to handle in LQC, even when they are
homogeneous.  This is because the holonomy of the Ashtekar connection
around a square loop in this case is not an almost periodic function of
the connection (see \cite{Ashtekar:2009um} for an explicit example of this
in the Bianchi II space-time).  If the holonomy around the square loop is
not almost periodic in the connection, then it is not known how to build
an operator corresponding to the field strength in LQC and then the
Hamiltonian cannot be defined by the introduction of a nonlocal field
strength operator as was done here.  This problem can be avoided by
introducing a nonlocal connection operator as in \cite{Ashtekar:2009um,
WilsonEwing:2010rh} in which case a nondiagonal metric can be handled,
thus making it possible to treat scalar, vector and tensor modes
together in a gauge-invariant LQC treatment; we leave this for future work.

A limitation of lattice LQC is that inverse triad effects cannot be included in
the model ---either in the quantum or the effective theories--- as one of the
necessary assumptions to work on the lattice in the first place was for each
cell to be larger than the Planck volume and therefore also for the wavelengths
of the modes of interest to remain greater than $\lp$, in which case inverse
triad corrections are expected to be negligible \cite{WilsonEwing:2011es}.
Inverse triad corrections could play an important role, particularly in
inflationary models where the wavelengths of the modes visible today were
much smaller than $\lp$ at the bounce point in LQC but then lattice LQC is
not an appropriate model to study such a scenario.  Nonetheless, we expect
lattice LQC to be an excellent model for cosmological scenarios where the
wavelengths of interest remain larger than the Planck length at all times,
including the ekpyrotic model and various matter bounce scenarios.

The main open question now is whether the quantum gravity effects studied here
leave some signal in the classical regime in cosmology and, if they do, how
they might be detected in for example the cosmic microwave background.  This
should be studied for the ekpyrotic, matter bounce and inflationary models (in
which case inverse triad effects could also be important), as well as all other
cosmological scenarios in order to precisely understand the observational
consequences of quantum gravity effects in each of these cosmologies, and
thereby test LQC.

\acknowledgments

The author would like to thank Martin Bojowald, Jerzy Lewandowski, Carlo Rovelli
and David Sloan for helpful discussions.
This work was supported by Le Fonds qu\'eb\'ecois de la recherche
sur la nature et les technologies.

%\bibliographystyle{utcaps}
%\bibliography{bibliography}
%\end{document}

\end{document}